\documentclass[fleqn,usenatbib]{mnras}

\usepackage{newtxtext,newtxmath}

\usepackage[T1]{fontenc}

\DeclareRobustCommand{\VAN}[3]{#2}
\let\VANthebibliography\thebibliography
\def\thebibliography{\DeclareRobustCommand{\VAN}[3]{##3}\VANthebibliography}

\usepackage{graphicx}	
\usepackage{amsmath}	
\usepackage{bm}
\usepackage{xspace}
\usepackage{xcolor}
\usepackage{tabularx}
\makeatletter 
  \patchcmd{\NAT@citex}
    {\@citea\NAT@hyper@{%
      \NAT@nmfmt{\NAT@nm}%
      \hyper@natlinkbreak{\NAT@aysep\NAT@spacechar}{\@citeb\@extra@b@citeb}%
      \NAT@date}}
    {\@citea\NAT@nmfmt{\NAT@nm}%
    \NAT@aysep\NAT@spacechar\NAT@hyper@{\NAT@date}}{}{}

  \patchcmd{\NAT@citex}
    {\@citea\NAT@hyper@{%
      \NAT@nmfmt{\NAT@nm}%
      \hyper@natlinkbreak{\NAT@spacechar\NAT@@open\if*#1*\else#1\NAT@spacechar\fi}%
        {\@citeb\@extra@b@citeb}%
      \NAT@date}}
    {\@citea\NAT@nmfmt{\NAT@nm}%
    \NAT@spacechar\NAT@@open\if*#1*\else#1\NAT@spacechar\fi\NAT@hyper@{\NAT@date}}
    {}{}
\makeatother

\newcommand{\msun}{\,\mathrm{M}_\odot}

\newcommand{\kms}{\,\mathrm{km}\,\mathrm{s}^{-1}}

\newcommand{\K}{\,\mathrm{K}}
\newcommand{\Myr}{\,\mathrm{Myr}}
\newcommand{\pc}{\,\mathrm{pc}}
\newcommand{\kpc}{\,\mathrm{kpc}}
\newcommand{\pkpc}{\,\mathrm{pkpc}}
\newcommand{\Mpc}{\,\mathrm{Mpc}}

\newcommand{\thesan}{\textsc{thesan}\xspace}
\newcommand{\thesanzoom}{\textsc{thesan-zoom}\xspace}

\definecolor{mycolor1}{HTML}{54278f}   
\definecolor{mycolor2}{HTML}{756bb1}   
\definecolor{mycolor3}{HTML}{9e9ac8}   
\definecolor{mycolor4}{HTML}{cbc9e2}   

\definecolor{mycolor5}{HTML}{7f2704}   
\definecolor{mycolor6}{HTML}{a63603}   
\definecolor{mycolor7}{HTML}{e6550d}   
\definecolor{mycolor8}{HTML}{fd8d3c}   
\definecolor{mycolor9}{HTML}{fdae6b}   
\definecolor{mycolor10}{HTML}{fee6ce}  

\def\aap{A\&A}

\def\apjl{ApJ}

\def\apjs{ApJS}

\title[Clumpiness at high-redshift]{The \thesanzoom project: clumpiness of high-redshift galaxies and its connection to bursty star formation}

\author[Z. Wang et al.]{\parbox{17.5cm}{
Zihao Wang$^{1,2,3}$\thanks{E-mail \href{mailto:zihaowang25@stu.pku.edu.cn}{zihaowang25@stu.pku.edu.cn}},
Xuejian Shen$^{2,4}$,
Rahul Kannan$^{5}$,
Ewald Puchwein$^{6}$,
Aaron Smith$^{7}$,
Josh Borrow$^{8}$,
Enrico Garaldi$^{9,10}$,
Laura Keating$^{11}$,
Mark Vogelsberger$^{2}$,
Oliver Zier$^{4}$,
William McClymont$^{12,13}$,
Sandro Tacchella$^{12,13}$,
Fangzhou Jiang$^{3}$,
Hui Li$^{14}$,
and
Lars Hernquist$^{4}$}
\\ \vspace{0.2cm} \\
$^1$ Department of Astronomy, Peking University, Beijing 100871, China \\
$^2$ Department of Physics, Kavli Institute for Astrophysics and Space Research, Massachusetts Institute of Technology, Cambridge, MA 02139, USA \\
$^3$ Kavli Institute for Astronomy and Astrophysics, Peking University, Beijing 100871, China\\
$^4$ Center for Astrophysics | Harvard \& Smithsonian, 60 Garden St, Cambridge, MA 02138, USA\\
$^5$ Department of Physics and Astronomy, York University, 4700 Keele Street, Toronto, ON M3J 1P3, Canada \\
$^6$ Leibniz-Institut f\"ur Astrophysik Potsdam, An der Sternwarte 16, 14482 Potsdam, Germany \\
$^7$ Department of Physics, The University of Texas at Dallas, Richardson, TX 75080, USA \\
$^8$ Department of Physics and Astronomy, University of Pennsylvania, 209 South 33rd Street, Philadelphia, PA 19104, USA \\
$^9$ Kavli Institute for the Physics and Mathematics of the Universe, The University of Tokyo, 5-1-5 Kashiwanoha, Kashiwa, 277-8583, Chiba, Japan \\
$^{10}$ Center for Data-Driven Discovery, Kavli IPMU (WPI), UTIAS, The University of Tokyo, Kashiwa, Chiba 277-8583, Japan \\
$^{11}$ Institute for Astronomy, University of Edinburgh, Blackford Hill, Edinburgh, EH9 3HJ, UK \\
$^{12}$ Kavli Institute for Cosmology, University of Cambridge, Madingley Road, Cambridge CB3 0HA, UK \\
$^{13}$ Cavendish Laboratory, University of Cambridge, 19 JJ Thomson Avenue, Cambridge CB3 0HE, UK \\
$^{14}$ Department of Astronomy, Tsinghua University, Beijing 100084, China \\
}

\date{Accepted XXX. Received YYY; in original form ZZZ}

\pubyear{2026}

\begin{document}
\label{firstpage}
\pagerange{\pageref{firstpage}-\pageref{lastpage}}
\maketitle

\begin{abstract}
Recent \textit{JWST} observations have revealed diverse high-redshift galaxy morphologies, including a population with irregular and clumpy structures. The physical origin of these structures, and the extent to which observational biases shape their appearance, remain uncertain. We present a power-spectrum-based method for quantifying galaxy clumpiness across spatial scales, using the radiation-hydrodynamic simulation suite \thesanzoom, which employs a state-of-the-art galaxy formation model that resolves the multiphase interstellar medium (ISM). Although the total stellar mass distributions in \thesanzoom galaxies are usually smooth, clumpy structures appear in the H$\alpha$, far-ultraviolet (FUV), and optical light distributions. Tracers sensitive to shorter-timescale star formation exhibit more pronounced small-scale structure ($\sim10^{2}$--$10^{3}\pc$). The corresponding projected light spectra follow $P(k)\propto k^{-1}$ to $k^{-2}$, with progressively shallower slopes for tracers sensitive to more recent star formation, reflecting enhanced small-scale power and greater spatial intermittency in young stellar populations. This behaviour is consistent with a highly compressible, shock-dominated ISM in which stellar feedback and outflows reorganise dense gas into filamentary and clumpy structures. We also find that galaxy clumpiness depends on the treatment of stellar feedback. Weaker early stellar feedback enhances small-scale power in both the mass and light distributions. Clumpiness also varies strongly over the bursty star formation cycle, implying that observed samples may be biased towards galaxies caught in phases of elevated star formation. Galaxy clumpiness, therefore, could provide a complementary probe of the bursty star formation in the early Universe.
\end{abstract}

\begin{keywords}
methods: numerical -- galaxies: high-redshift -- galaxies: structure
\end{keywords}



\section{Introduction}
The morphological evolution of galaxies across cosmic time provides an important window into galaxy formation and evolution. \textit{Hubble Space Telescope} (\textit{HST}) imaging shows that galaxies at cosmic noon ($0.5<z<3$) often have irregular structures, characterised by patchy, clumpy substructure on kiloparsec scales \citep[e.g.,][]{Cowie1995,Giavalisco1996,vandenBergh1996,Elmegreen2005,Elmegreen2007,Elmegreen2009,Bournaud2007,Forster2011,Sattari2023}. These clumps are bright in the rest-frame ultraviolet (UV), relatively massive ($\gtrsim10^7\msun$), and typically young, with stellar ages of $\sim100\Myr$. On average, they contribute only about 4-20\% of the total stellar mass but can account for up to 50\% of their host galaxy's total star formation rate \citep[SFR;][]{Elmegreen2008,Elmegreen2009,Forster2011,Guo2012,Guo2015,Wisnioski2012,Wuyts2012,Wuyts2013,Livermore2015,Shibuya2016,Soto2017,Zanella2019,Mehta2021,Vanzella2021,Mestric2022}.
These features have long been interpreted as signatures of distinct growth modes of galaxies at high redshift, motivating theoretical pictures in which star formation is driven by violent disc instabilities \citep[e.g.,][]{Noguchi1998,Noguchi1999,Immeli2004,Immeli2004a,Bournaud2007,Bournaud2009,Agertz2009,Dekel2009,Ceverino2010,Romeo2010,Romeo2014,Inoue2016,Meng2019,Orr2024}, or frequent mergers \citep[e.g.,][]{DiMatteo2008,Renaud2015,Li2017,Li2018,Li2022,Nakazato2024,Deng2025}.

The advent of the \textit{James Webb Space Telescope} (\textit{JWST}) has enabled these questions to be revisited with greater sensitivity and spatial resolution, extending such studies to earlier cosmic epochs at $z>3$ and revealing a more complex picture. On the one hand, \textit{JWST} imaging over $1<z<8.5$ has uncovered a substantial population of clumps that were not detected with \textit{HST} \citep[e.g.,][]{Claeyssens2023,Tacchella2023,Hainline2024,Vega2026,Kalita2025,Kalita2025b,Mercier2026}, and observations of some strongly lensed systems even resolve dense star clusters~\citep[e.g.,][]{Vanzella2023,Adamo2024,Messa2024,Mowla2024,Bradley2025,Fujimoto2025}. Early studies also report a higher fraction of clumpy galaxies \citep{Vega2026} and a larger contribution from clumps to the total UV light ($\gtrsim70$\%; \citealt{Chen2023}) than the $\sim20$-30\% measured over $0.5<z<3$ \citep{Guo2015}. On the other hand, \textit{JWST} has uncovered surprisingly regular, extended, disc-like galaxies at similar redshifts \citep[e.g.,][]{Kartaltepe2023,Costantin2023,Vegaferrero2023,Lee2024,Wang2025}. Complementary ALMA observations of bright, lensed galaxies have also detected dynamically cold, rotating gaseous discs \citep[e.g.,][]{Rizzo2020,Jones2021,RO2023,Fujimoto2025,Rowland2024} with high rotation-to-dispersion ratios ($V_{\mathrm{rot}}/\sigma\gtrsim10$), in some cases coexisting with prominent clumps \citep{Fujimoto2025}. These findings suggest the early emergence of mature galaxy structures, challenging the traditional picture in which high-redshift galaxies are predominantly irregular and clumpy \citep[e.g.,][]{Shen2024b-ede,Shen2026,Semenov2025}.

A potential caveat in interpreting these observations is that apparent morphology may depend strongly on sample selection and observational tracer. The galaxies most readily detected at high redshift may be biased towards phases of elevated star formation \citep[e.g.,][]{Madau2014,Donnan2023,Harikane2023}, during which their apparent morphologies may differ systematically from those of galaxies on the star-forming main sequence. A related issue is the connection between morphologies inferred in different wavelength bands and the underlying stellar mass distribution. Previous studies have shown that both the identification and measured properties of clumps depend sensitively on the tracer and wavelength used \citep[e.g.,][]{Wuyts2012,Buck2017,Mager2018}. At a given redshift, \textit{JWST} extends morphological studies to longer rest-frame wavelengths than \textit{HST}, thereby tracing different stellar populations and exposing complementary aspects of galaxy structure \citep[e.g.,][]{Kalita2025b,Mercier2026}. Differences among these tracers reflect the evolution of stellar populations after formation and provide valuable insights into how galaxies assemble and evolve over cosmic time. Establishing the connection between mass and light is therefore essential for interpreting the morphological evolution of galaxies.

Beyond their observational interest, galaxy morphologies provide a powerful test of galaxy formation models. Physical processes, including gas accretion, star formation, and stellar feedback, leave distinct imprints on the spatial distributions of stars and gas \citep[e.g.,][]{Dekel2009,Ceverino2010,Hopkins2014,Ma2018,Shen2024c-size}. In particular, the abundance and properties of clumpy structures are expected to depend sensitively on feedback strength and star formation efficiency \citep[e.g.,][]{Bournaud2014,Oklop2016,Mandelker2017,Ceverino2023,Shin2023}. Comparing morphological statistics from simulations and observations, therefore, offers a route to constraining the processes that shape galaxy evolution.

In this work, we use the \thesanzoom simulation suite \citep{Kannan2025} to investigate these issues by comparing the spatial structure of galaxies in stellar mass and light. The \thesanzoom campaign is designed to provide realistic theoretical counterparts to the diverse high-redshift galaxy population revealed by \textit{JWST}. These zoom-in radiation-hydrodynamic simulations resolve the multiphase interstellar medium (ISM) at high spatial and mass resolution at $z\gtrsim3$, enabling detailed studies of internal galaxy structure and star formation. The suite has already produced a number of early results on high-redshift galaxy formation and evolution \citep[e.g.,][]{McClymont2025-MsScatter,McClymont2025-sm,McClymont2025-Metal,Zier2025-reion,Zier2025-PopIII,Shen2025-SFE,Wang2025-SFE,Pruto2026,Summerfield2026}, demonstrating its ability to yield galaxy populations broadly consistent with current observational constraints. Building on these capabilities, we quantify galaxy morphology using the power spectra of intrinsic stellar mass density fields and mock emission maps, providing a scale-dependent characterisation of the connections and discrepancies between the mass and light distributions of simulated high-redshift galaxies.

This paper is organised as follows. In Section~\ref{sec:methods}, we introduce the simulation suite and analysis methods. In Section~\ref{sec:results}, we present the relation between the power spectra of intrinsic mass density fields and mock emission maps. In Section~\ref{sec:discussions}, we assess several observational effects, explore physical and numerical variations in the galaxy formation model, and examine the connection between clumpiness and star formation activity. Throughout, we adopt the cosmological parameters inferred by \citet{Planck2016} from their TT,TE,EE+lowP+lensing+BAO+JLA+H0 data set. These are $H_0=67.74\,\kms/\Mpc$, $\Omega_{\mathrm{m}}=0.3089$, $\Omega_{\Lambda}=0.6911$, $\Omega_{\mathrm{b}}=0.0486$, $\sigma_8=0.8159$, and $n_{\mathrm{s}}=0.9667$.

\section{Methods}\label{sec:methods}
\subsection{Simulations}
\begin{table}
    \centering
    \caption{Numerical parameters of the \thesanzoom simulation suite. From left to right, the columns give the following quantities. \\
    (1) the resolution level. All target haloes were simulated at the ``$4\times$'' resolution level, whereas only the low-mass haloes were simulated at ``$8\times$'' and ``$16\times$''; \\
    (2) the effective (total-volume-equivalent) number of particles; \\
    (3,4) the masses of high-resolution dark matter (DM) particles and gas cells in the initial conditions; \\
    (5) the Plummer-equivalent comoving softening length of star and DM particles; and \\
    (6) the minimum comoving softening length of gas cells.}
    \label{table:res}
    \addtolength{\tabcolsep}{-0.2pt}
    \def\arraystretch{1.2}
    \begin{tabular}{lccccc} 
	\hline
	Name & $N_{\mathrm{part}}^{\mathrm{eff}}$ & $m_{\mathrm{DM}}$ & $m_{\mathrm{gas}}$ & $\epsilon_{\mathrm{DM},\ast}$ & $\epsilon_{\mathrm{gas}}^{\mathrm{min}}$\\
		& & [$\mathrm{M}_\odot$] & [$\mathrm{M}_\odot$] & [cpc] & [cpc]\\
		\hline
            $4\times$ & $2 \times 8400^3$ & $4.86 \times 10^4$ & $9.09 \times 10^3$ & $553.59$ & $69.20$\\
            $8\times$ & $2 \times 16\,800^3$ & $6.09 \times 10^3$ & $1.14 \times 10^3$ & $276.79$ & $34.60$\\
            $16\times$ & $2 \times 33\,600^3$  & $7.62 \times 10^2$ & $ 1.42 \times 10^2$ & $138.30$ & $17.30$\\
		\hline
	\end{tabular}
\end{table}
We use the high-resolution, zoom-in \thesanzoom simulations. A full description is given by \citet{Kannan2025}; here we summarise their key features. The \thesanzoom runs were performed with \textsc{arepo-rt} \citep{Kannan2019}, a radiation-hydrodynamic extension of the moving-mesh code \textsc{arepo} \citep{Springel2010}. They adopt a state-of-the-art galaxy formation model that resolves the multiphase interstellar medium \citep[ISM;][]{Marinacci2019,Kannan2020,Zier2024}. Resolving the ISM allows us to study internal galaxy processes that large-volume cosmological simulations typically leave unresolved or approximate using effective equation-of-state prescriptions \citep{Springel2003a,Pillepich2018}. Table~\ref{table:res} lists the numerical parameters used at each resolution level.

The simulations use a zoom-in technique to resimulate selected regions of the \thesan parent volume \citep{Kannan2022thesan,Smith2022,Garaldi2022,Garaldi2024}. The time-dependent radiation field from the parent box is injected at the boundaries of the zoom-in region during inflow, allowing radiative feedback from nearby galaxies to affect the \thesanzoom targets.
Halo catalogues are constructed with the friends-of-friends (FoF) algorithm \citep{Davis1985}, and gravitationally bound subhaloes within FoF groups are identified with the \textsc{SUBFIND-HBT} algorithm \citep{Springel2021}. In this work, we focus exclusively on a subset of the most massive main-target galaxies in the $4\times$-resolution suite, which has a median baryonic mass resolution of $9.09\times10^3\msun$. We assess the numerical convergence of our results separately in Appendix~\ref{apdx:res}. The selected galaxies are ``m13.0'', ``m12.6'', ``m12.2'', ``m11.9'', ``m11.5'', ``m11.1'', ``m10.8'', and ``m10.4''. Of these, ``m13.0'' is evolved only to $z\simeq6$, while the others are run to $z=3$. Their stellar masses are broadly consistent with the observed relation between stellar mass and halo mass \citep{Kannan2025,Shen2025-SFE}.

\begin{figure}
    \centering
    \includegraphics[width=0.95\linewidth]{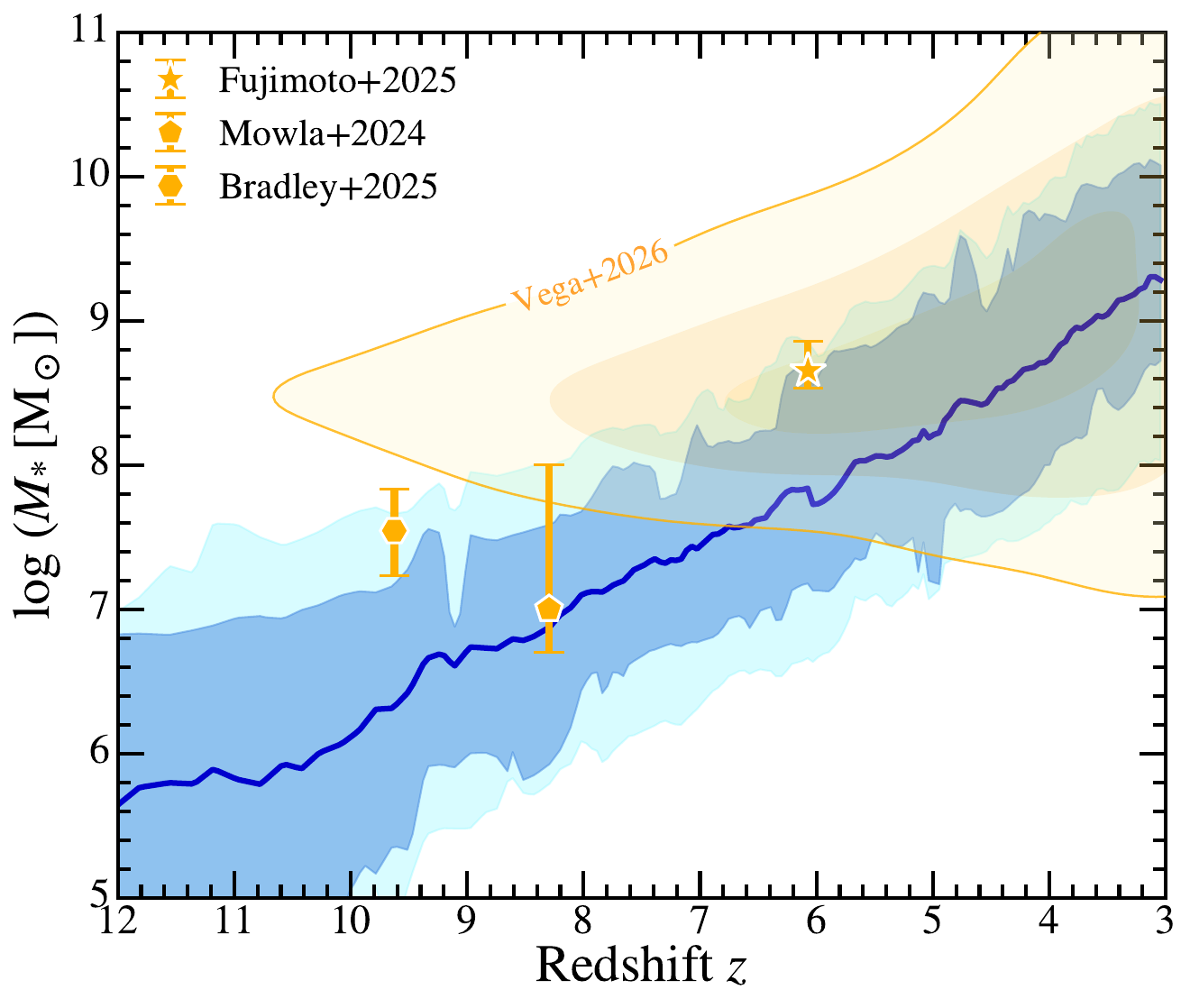}
    \caption{Stellar masses of the galaxy sample as a function of redshift. Each stellar mass is the total enclosed within $R_{\mathrm{gal}}=\min(2R_{1/2,\ast},0.15R_{\mathrm{vir}})$, where $R_{1/2,\ast}$ is the stellar half-mass radius. The solid line shows the median stellar mass of the \thesanzoom target galaxies, and the shaded regions denote the $1\sigma$ and $2\sigma$ ranges. For observational context, we also show recent \textit{JWST} measurements of clumpy galaxies \citep{Vega2026} and strongly lensed systems \citep[e.g.,][]{Fujimoto2025,Mowla2024,Bradley2025}.}
    \label{fig:galaxy}
\end{figure}

\begin{figure*}
    \centering
    \includegraphics[width=1\linewidth]{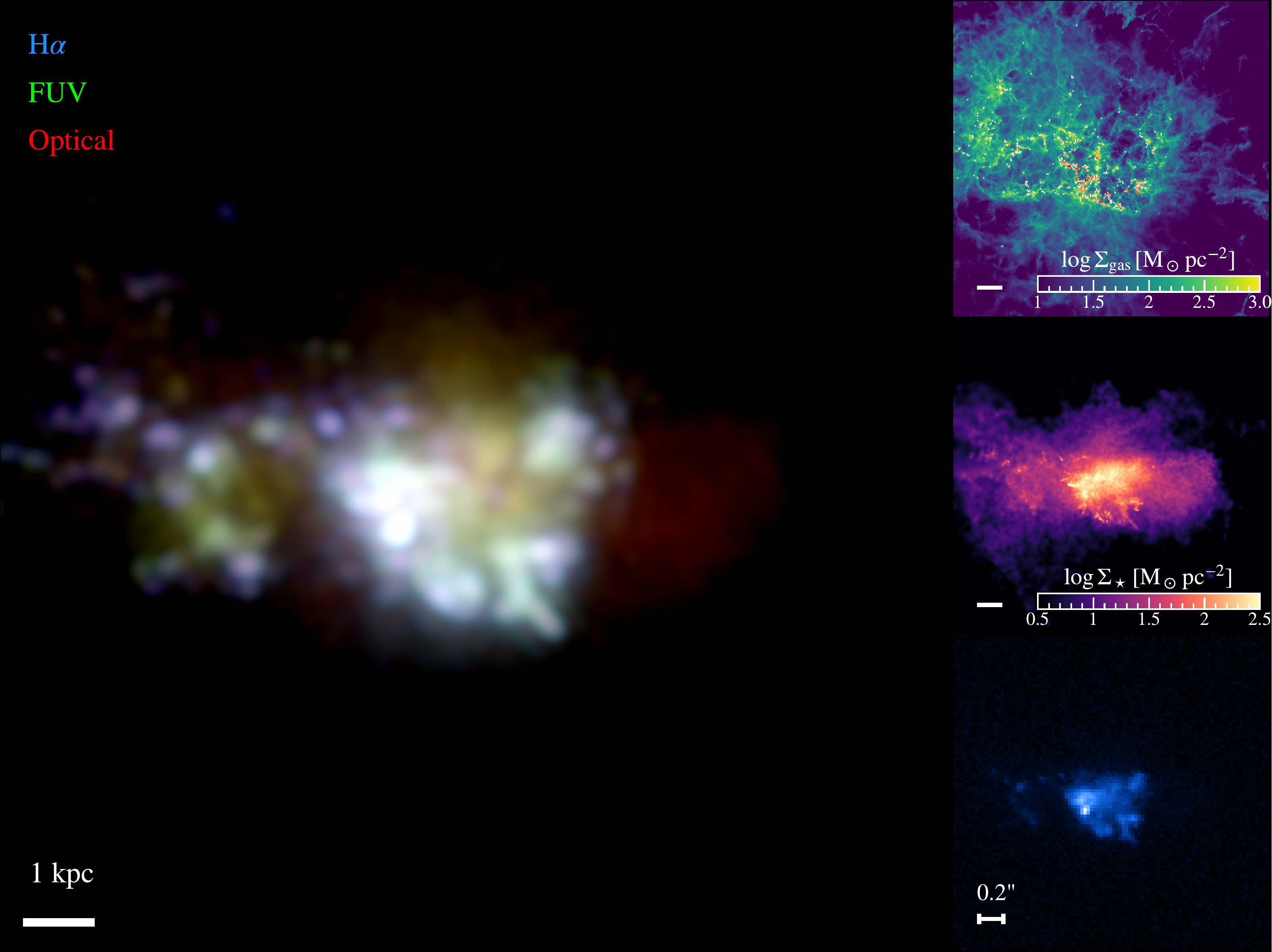}
    \caption{Visualisation of the \thesanzoom galaxy ``m12.6'' at $z\simeq6$ within a $15\kpc$ field of view. The main panels show RGB composites in which colour encodes the logarithmic overdensity of H$\alpha$ (blue), rest-frame FUV (green), and optical emission (red), computed following Equation~\eqref{eq:od}. For a consistent visual comparison, all three fields are smoothed with the same \textit{JWST}/NIRCam F115W point spread function (PSF).
    From top to bottom, the right-hand panels show the gas surface density overlaid with the instantaneous SFR surface density, the stellar surface density, and example mock FUV images including PSF convolution and noise.
    The stellar mass distribution differs markedly from the emission maps. Even when the stellar mass distribution is relatively smooth, tracers of recent star formation remain strongly clumpy.}
    \label{fig:vis}
\end{figure*}

\subsection{Galaxy identification and physical properties}
We focus on a subset of the most massive main-target central galaxies in the simulations. This choice is motivated by two considerations. Current observations primarily probe this mass regime, and more massive hosts tend to contain stellar clumps that are better resolved in the simulations.

We define the boundary of each galaxy by $R_{\mathrm{gal}}=\min(2R_{1/2,\ast},0.15R_{\mathrm{vir}})$, where $R_{1/2,\ast}$ is the stellar half-mass radius. The upper limit of $0.15R_{\mathrm{vir}}$ accounts for the more extended nature of high-redshift galaxies. Using only the stellar half-mass radius may include satellite galaxies or nearby substructures~\citep[e.g.,][]{Liang2026}, thereby biasing the clumpiness measurement.
Throughout this paper, all galaxy properties, including stellar mass $M_\ast$ and SFR, are computed within a spherical region of radius $R_{\mathrm{gal}}$ centered on the particle with the minimum gravitational potential. Figure~\ref{fig:galaxy} shows the stellar masses of our sample as a function of redshift. The mass range broadly overlaps that probed by \textit{JWST} observations of clumpy galaxies \citep[e.g.,][]{Fujimoto2025,Mowla2024,Bradley2025,Vega2026}.

\subsection{Power spectrum measurement}\label{subsec:PS}

Motivated by non-parametric clumpiness measures \citep[e.g.,][]{Conselice2003}, we describe galaxy clumpiness through the scale-dependent fluctuations in the stellar-mass or light distributions. These fluctuations are captured by the power spectrum of the corresponding density or emission field \citep{Willett2005,Shin2023}, allowing clumpiness to be quantified as a function of physical scale rather than compressed into a single number.
For a density or surface-brightness field $\rho(\bm{r})$, we first compute the overdensity
\begin{equation}
\label{eq:od}
\delta(\bm{r}) = \frac{\rho(\bm{r})}{\langle\rho\rangle}-1\, ,
\end{equation}
where $\bm{r}$ is the position of each cell (or pixel), and $\langle\rho\rangle$ is the mean field value across the map or volume. We then compute the discrete Fourier transform of the overdensity. For a $d$-dimensional grid of side length $L$ with $N$ cells per dimension, the allowed Fourier modes are
\begin{equation}
\bm{k}_{\bm{n}} = \frac{2\pi}{L}\bm{n}\, ,
\end{equation}
where $\bm{n}$ is an integer vector. The Fourier amplitudes and isotropically averaged power spectrum are
\begin{equation}
\begin{split}
\tilde{\delta}_{d}(\bm{k}_{\bm{n}})
&=
\sum_{\bm{j}}
\delta(\bm{r}_{\bm{j}})
\exp\left(-\mathrm{i}\bm{k}_{\bm{n}}\cdot\bm{r}_{\bm{j}}\right),
\\[3pt]
P_{d}(k)
&=
\frac{L^{d}}{N^{2d}}
\frac{1}{N_{k}}
\sum_{k\leq|\bm{k}_{\bm{n}}|<k+\Delta k}
\left|\tilde{\delta}_{d}(\bm{k}_{\bm{n}})\right|^{2}\, .
\end{split}
\label{eq:FFT}
\end{equation}
Here, $\bm{j}$ labels grid cells, and $N_k$ is the number of modes in the shell for $d=3$ or the annulus for $d=2$.

In practice, we extract a cubic region of side length $L=2R_{\mathrm{gal}}$ centred on each target galaxy. We assign stellar particles or gas cells to a uniform Cartesian grid using a particle-in-cell (PIC) scheme and compute the Fourier transform of the resulting overdensity field. Unless otherwise stated, we adopt a spatial resolution of $\Delta x=50\pc$, close to the finest resolution that the stellar particle distribution can robustly sample. We construct mock images and projected surface-density maps analogously, projecting the particle emission or mass along the line of sight onto a two-dimensional Cartesian grid with the same spatial resolution.

\subsection{Mock observations}\label{subsec:mock}
We generate emission maps of the simulated galaxies using \textsc{synthesizer}\footnote{\url{https://synthesizer-project.github.io/synthesizer/index.html}} \citep{Lovell2025,Roper2026}. Specifically, we adopt the BPASS v2.2.1 stellar spectral templates, which include the effects of binary evolution \citep{Eldridge2017,Stanway2018}, to construct the rest-frame UV and optical spectral energy distributions (SEDs). We assume the \citet{Chabrier2003} initial mass function (IMF) over a mass range of $0.1$--$300\msun$.
The spectra have been post-processed with the photoionisation code \textsc{Cloudy} v23.01 \citep{Ferland1998,Chatzikos2023} to compute the associated nebular continuum and line emission. We adopt a fiducial hydrogen density of $n_{\mathrm{H}}=1000\,\mathrm{cm}^{-3}$ and an ionisation parameter of $U=0.01$, consistent with values inferred for high-redshift galaxies from recent \textit{JWST} observations \citep[e.g.,][]{Wilkins2020,Reddy2023,Calabro2024,Topping2024}.
For each star particle, we evaluate the spectral emissivity from its age and initial metallicity. We then construct mock images by summing the particle emission along the line of sight in the PIC scheme, weighted by initial stellar mass, at a default spatial resolution of $\Delta x=50\pc$. Spatially resolved H$\alpha$, rest-frame UV, and optical fluxes are computed by extracting the line emission and applying top-hat filters. Specifically, we extract H$\alpha$ emission at $\lambda=6562.8$~\AA, while the UV and optical continuum bands use top-hat filters spanning 1450--1550~\AA\ and 5070--5950~\AA, respectively. Figure~\ref{fig:vis} compares these tracers for the representative \thesanzoom galaxy ``m12.6'' at $z\simeq6$. Their emission maps show different degrees of clumpiness and differ markedly from the smoother stellar mass distribution.

For simplicity, we neglect dust attenuation. Our primary goal is to understand the intrinsic connection between stellar mass distributions and emission-based tracers of recent star formation, for which we expect the qualitative trends to be robust to moderate dust obscuration. A self-consistent treatment of both dust attenuation and re-emission would require additional radiative-transfer modelling and calibrations beyond the scope of our current mock-observation framework. This simplification is also reasonable for the physical regime considered here. High-redshift galaxies with stellar masses and UV luminosities comparable to our sample ($M_{\mathrm{UV}}\gtrsim-22$ at $z\gtrsim5$) generally have blue observed UV continuum slopes, implying only modest dust attenuation, with $A_{1500}\lesssim1$ \citep[e.g.,][]{Ma2018,Shen2020}.


\section{From mass clumps to light clumps}\label{sec:results}

\subsection{Intrinsic stellar mass distribution}
\begin{figure*}
    \centering
    \includegraphics[width=1\linewidth]{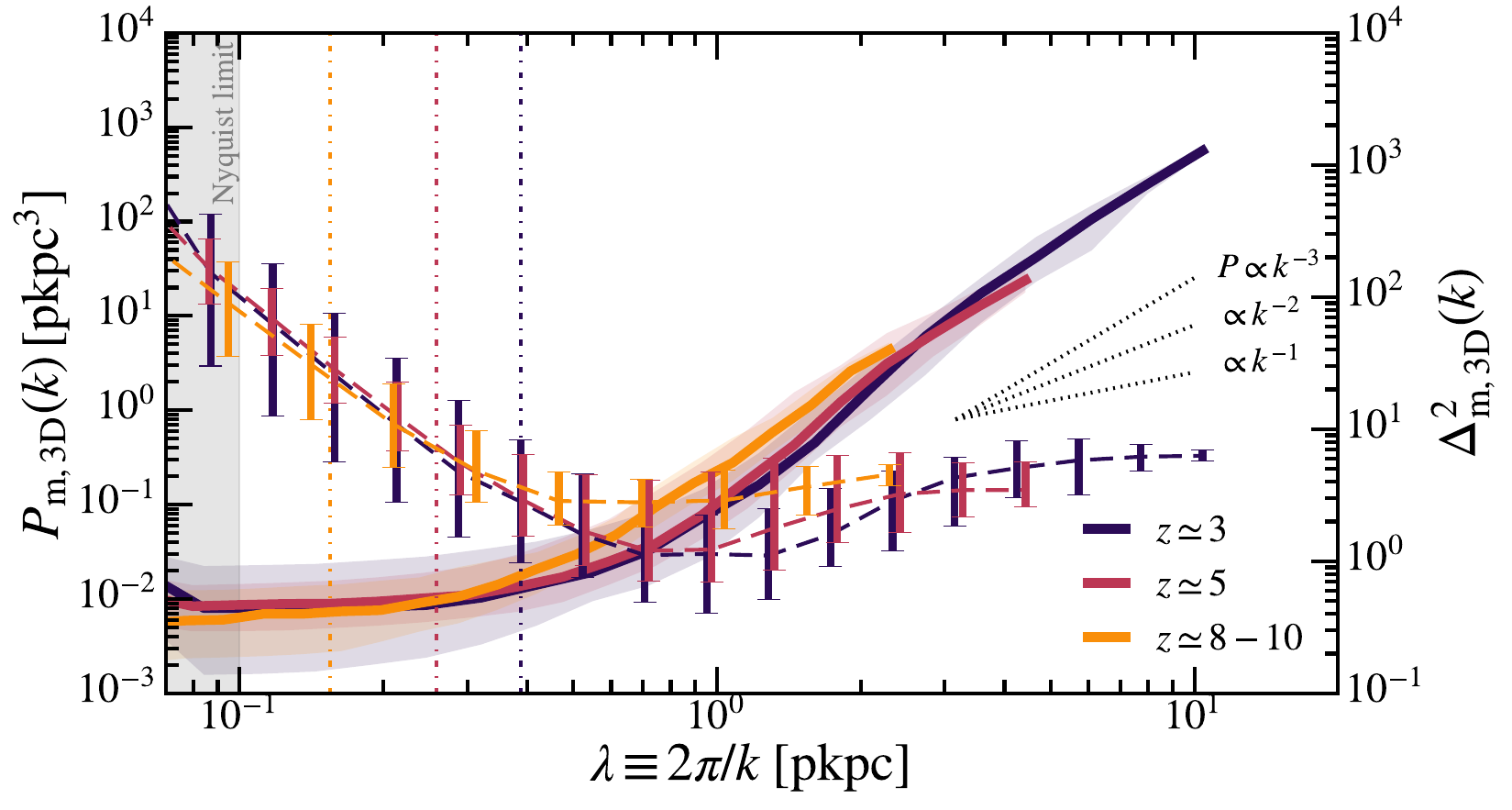}
    \caption{Median three-dimensional stellar mass density power spectra (solid) and corresponding dimensionless power spectra (dashed), $\Delta_{\mathrm{3D}}^2(k)=k^3P_{\mathrm{3D}}(k)/(2\pi^2)$, for galaxies from $z\simeq8$--$10$ to $z\simeq3$. Shaded regions and error bars denote the $16^{\mathrm{th}}$--$84^{\mathrm{th}}$ percentile variation across galaxies and snapshots. The wavenumber $k$ is converted to spatial scale using $\lambda\equiv2\pi/k$. The grey region denotes the Nyquist scale, $2\Delta x$, and vertical dashed lines mark the radius of compact support, $2.8\epsilon_\ast$, beyond which the gravitational force becomes Newtonian. The \thesanzoom galaxies have smooth, approximately power-law spectra with no prominent preferred Fourier scale above the numerical floor.}
    \label{fig:FFT}
\end{figure*}

We begin in Figure~\ref{fig:FFT} with the power spectra of the stellar mass distributions, which we refer to as the \textit{intrinsic} power spectra. We consider simulated galaxies at $z\simeq3$, $5$, and $8$--$10$. In each redshift bin, we select snapshots within a time interval $t_{\mathrm{dyn}}\simeq t_{\mathrm{H}}(z)/10$, where $t_{\mathrm{H}}(z)\equiv 1/H(z)$ is the Hubble time. To examine the redshift dependence, we take the median across all galaxies and snapshots in a given bin as its representative spectrum. Each galaxy is given equal weight with snapshots sampled at roughly $10\Myr$.
Appendix~\ref{apdx:ind} presents the spectra of individual galaxies and snapshots, including their short-timescale variation. For a more intuitive interpretation, we convert $k$ to its corresponding spatial scale using $\lambda\equiv2\pi/k$.
Dashed lines show the dimensionless spectra, $\Delta_{\mathrm{3D}}^2(k)=k^3P_{\mathrm{3D}}(k)/(2\pi^2)$, which quantify the contribution to the total stellar overdensity variance per ${\rm d}\ln k$.

The power spectra approximately follow a power law on large scales ($>1\pkpc$). Their overall form agrees with that found in idealised simulations of Milky Way-mass galaxies \citep[e.g.,][]{Shin2023}, showing that the stellar mass distribution is smooth and lacks a preferred clump scale over this range. Towards smaller scales, the spectra approach a numerical floor, where a white-noise-like signal becomes apparent. The scale at which this floor appears is roughly consistent with the radius of compact support of gravitational softening kernels of stellar particles ($2.8\times\epsilon_\ast$) in the simulations, beyond which gravity becomes exactly Newtonian. Since $\epsilon_\ast$ is set constant in comoving units in \thesanzoom, it corresponds to a redshift-dependent physical scale. Appendix~\ref{apdx:res} demonstrates this behaviour further by comparing spectra at different numerical resolutions. The overall smoothness of the spectra shows no preferred structural scale, although transient clumpy structures can still introduce short-timescale fluctuations in individual snapshots. The spectra evolve little with redshift apart from extending to larger spatial scales, consistent with the continued growth of either the galaxy as a whole or its central smooth component.

\subsection{Projected mass distribution}
\begin{figure}
    \centering
    \includegraphics[width=0.95\linewidth]{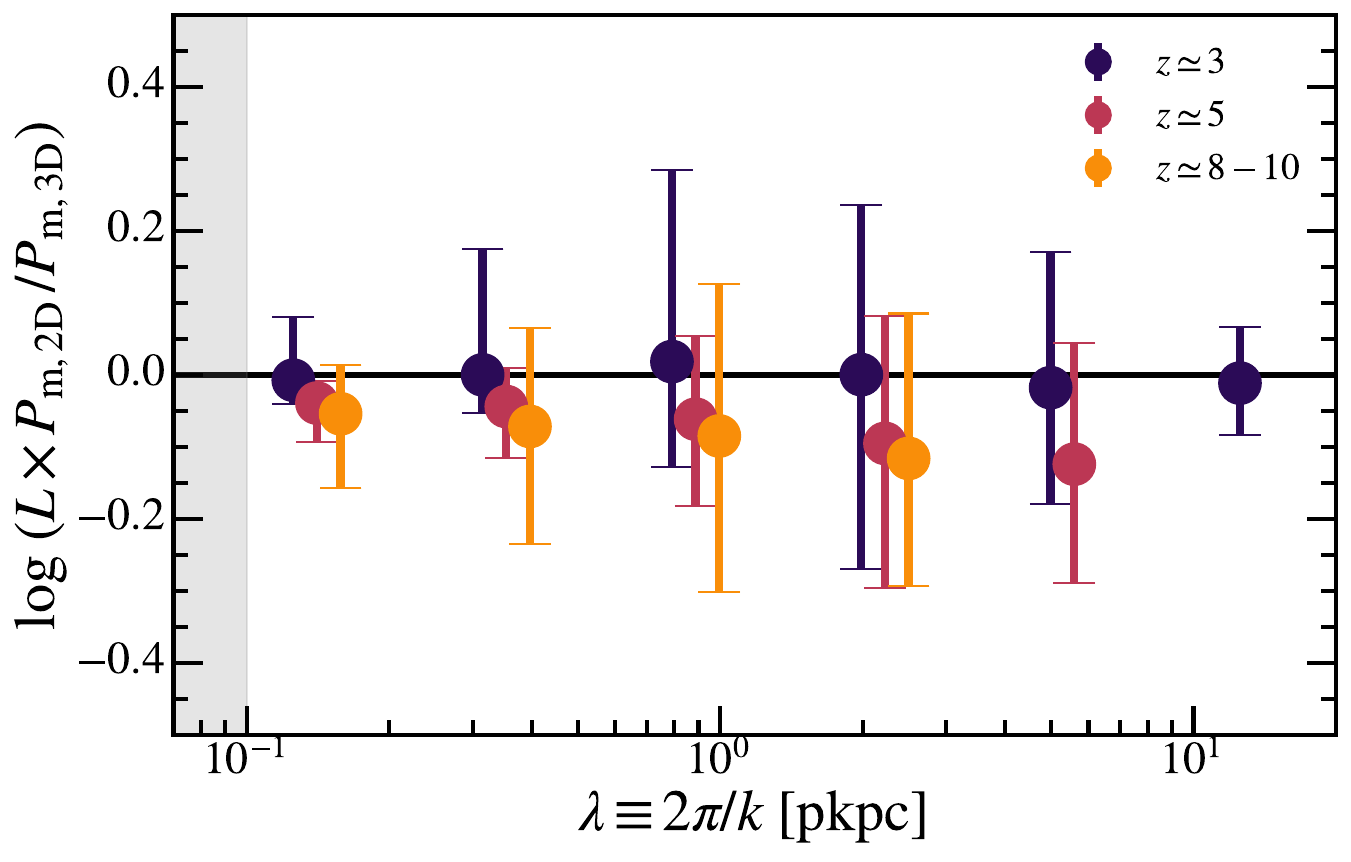}

    \caption{Ratio of projected to three-dimensional power spectra for \thesanzoom galaxies. Points show the median relation, and error bars denote the $1\sigma$ scatter. A ratio close to unity means that the projection preserves the underlying three-dimensional spectrum. At higher redshift, the projected spectra are systematically lower by less than $0.1\,\mathrm{dex}$, consistent with line-of-sight averaging slightly smoothing the more irregular structures in these systems. The scatter peaks at intermediate scales, where projection-induced stochasticity is strongest, at $\sim1\pkpc$.}
    \label{fig:FFT-2D}
\end{figure}

We next examine how projection affects the power spectra, providing a natural bridge between the intrinsic three-dimensional distribution and our analysis of two-dimensional emission maps. We project each galaxy along 10 random lines of sight and compute the corresponding two-dimensional power spectrum for every viewing direction.

For a statistically isotropic field, the projected two-dimensional power spectrum is related to its three-dimensional counterpart through an integration over line-of-sight Fourier modes:
\begin{equation}
P_{\mathrm{2D}}(k_\perp)=\frac{1}{2\pi}
\int_{-\infty}^{\infty}
\mathrm{d}k_z\,P_{\mathrm{3D}}
\left(\sqrt{k_\perp^2+k_z^2}\right)
\mathrm{sinc}^2\left(\frac{k_z L}{2}\right),
\end{equation}
where the $\mathrm{sinc}^2(k_z L/2)$ term arises from the finite top-hat projection of depth $L$. The two-dimensional spectrum is therefore a depth-averaged version of the three-dimensional spectrum rather than an independent quantity. If $P_{\mathrm{3D}}$ varies slowly over the line-of-sight modes admitted by the projection window, the integral reduces approximately to $P_{\mathrm{2D}}\sim P_{\mathrm{3D}}/L$. Under this assumption, the appropriate dimensionally matched comparison is between $L P_{\mathrm{2D}}$ and $P_{\mathrm{3D}}$.

Figure~\ref{fig:FFT-2D} shows that the projected power spectra broadly follow the isotropic expectation, meaning that projection largely preserves the underlying three-dimensional distribution. At higher redshifts, they show a mild systematic suppression of less than $0.1\,\mathrm{dex}$, consistent with line-of-sight averaging slightly smoothing the intrinsically more irregular and clumpy structures in these systems. The line-of-sight scatter quantifies the uncertainty introduced by viewing the same galaxy from different directions. This projection-induced uncertainty depends on scale. It is small on the largest scales, where the overall extent of the galaxy is preserved, and also near the resolution limit, where little coherent structure remains. The scatter is largest at intermediate, approximately kiloparsec scales.

\subsection{Projected light distribution}
We next examine the power spectra of emission fields, which are directly connected to observational measurements, focusing on rest-frame FUV, optical, and H$\alpha$ emission. We generate mock images along random lines of sight following Section~\ref{subsec:mock} and compute their power spectra using the method in Section~\ref{subsec:PS}.

\begin{figure*}
    \includegraphics[width=1\linewidth]{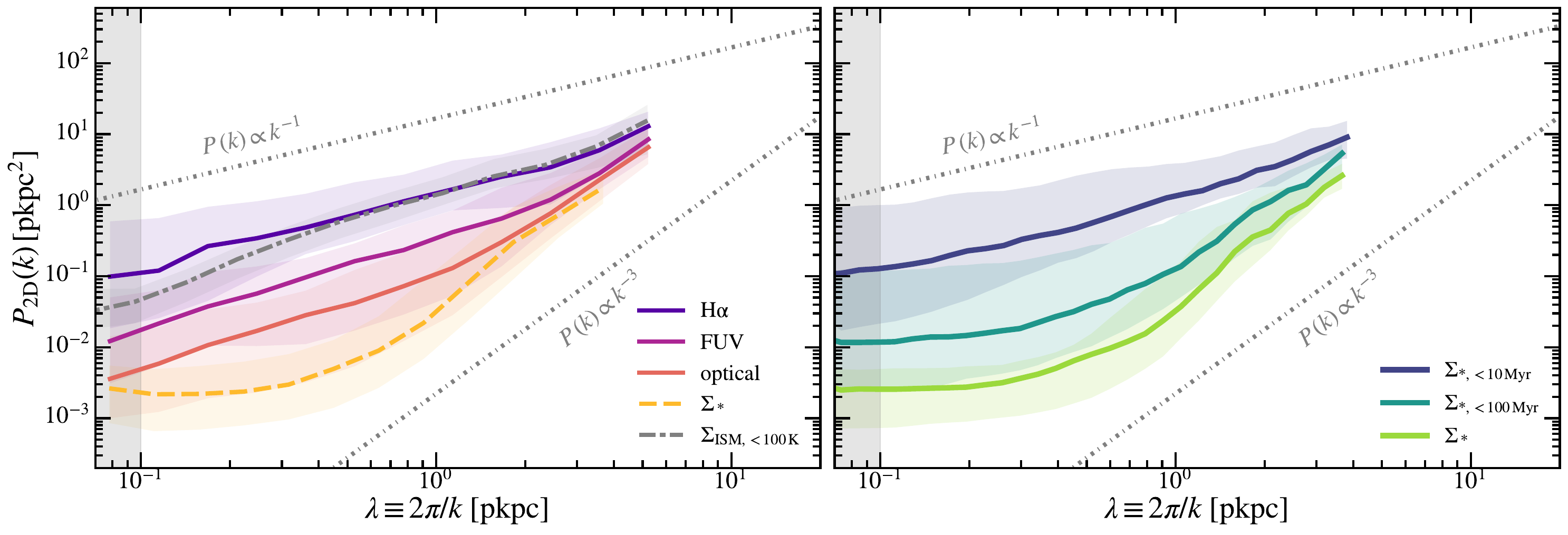}
    \caption{
    \textit{Left:} Median projected power spectra of rest-frame FUV, optical, and H$\alpha$ emission, compared with those of the stellar surface density and cold ISM ($T<10^{2}\K$), for \thesanzoom galaxies at $z\simeq6$. Shaded regions denote the $16^{\mathrm{th}}$--$84^{\mathrm{th}}$ percentile variation across galaxies and snapshots. The cold-ISM curve is vertically rescaled; only its slope is compared. On large scales, all tracers approach the stellar mass distribution. Towards smaller scales, tracers more sensitive to recent star formation have progressively higher amplitudes and flatter power-law slopes, revealing enhanced small-scale structure. \textit{Right:} As in the left panel, but for stellar surface density maps constructed using cumulative upper-age cuts ($<10 \Myr$, $<100 \Myr$, and all ages). Younger populations have systematically flatter power spectra and enhanced small-scale power, and converge towards the total stellar mass distribution with increasing age.
    }
    \label{fig:light}
\end{figure*}

The left panel of Figure~\ref{fig:light} presents the median power spectrum of each tracer together with the median projected stellar mass spectrum for reference. At low frequencies, corresponding to large scales, all tracers approach the stellar mass distribution, showing that the smooth large-scale component of the galaxies is largely insensitive to tracer choice. At the smallest scales, however, the light spectra lack the plateau seen in the mass density spectrum. This difference arises in part from the distinct numerical treatments of gas cells and stellar particles in the simulations. Gas cells have adaptive and generally smaller gravitational softening lengths. Tracers associated with newly formed stars can therefore retain small-scale structure inherited directly from the gas, even when this structure is smoothed in the stellar particle distribution. 
At intermediate scales ($0.1$--$1\kpc$), the spectra approximately follow a power law $P(k)\propto k^\alpha$. Tracers that are more sensitive to recent star formation have progressively shallower slopes, with H$\alpha$ giving the flattest spectrum, $\alpha\simeq-1$.

As discussed extensively in studies of turbulent ISM density fields \citep[e.g.,][]{Lazarian1995,Goldman2000,Elmegreen2001,Beresnyak2005,Dutta2013}, the spectral slope encodes the distribution of structure across scales and, in particular, the competition between pressure support and velocity fluctuations. In the incompressible, subsonic flows described by the classic Kolmogorov theory \citep{Kolmogorov1941}, kinetic energy cascades self-similarly from large to small scales, and the shell-integrated velocity spectrum follows $E(k)\propto k^{-5/3}$. Under the usual isotropic convention, this corresponds to a three-dimensional power spectrum $P(k)\propto k^{-11/3}$ \citep{FGD2014}. In this regime, pressure forces efficiently smooth strong compressions and prevent highly localised density enhancements from forming.

However, the ISM usually lies in a different regime. As the flow becomes increasingly supersonic and compressible, shocks and nonlinear compression begin to dominate the dynamics \citep{Vestuto2003}. Rather than producing a smooth, self-similar cascade, these processes reorganise the gas into a network of dense sheets, filaments, and compact clumps. In this regime, the density power spectrum becomes progressively shallower because a substantial fraction of the variance is concentrated in localised structures instead of being distributed evenly across scales \citep[e.g.,][]{Moraghan2015,Konstandin2016}.
A shallow spectral slope can therefore mark a highly compressible and intermittent medium. This behaviour resembles Burgers-like, shock-dominated turbulence \citep{Burgers1948}, whose dynamics are governed by discontinuities and interacting shocks. Simulations of feedback-driven turbulence likewise produce shallow density spectra with $\alpha\simeq-1$. In these simulations, expanding shells and cavities continually reshape the medium and concentrate power in small-scale structures \citep[e.g.,][]{Moraghan2015}.

Because H$\alpha$ traces star formation over $\sim10\Myr$, its shallow slope implies that high-redshift star formation is concentrated in compact, short-lived, and strongly clustered sites. In this picture, stars form preferentially in dense, shock-compressed filaments and knots assembled by supersonic turbulence and feedback. This interpretation is consistent with star formation occurring predominantly in filamentary giant molecular clouds (GMCs) embedded in a dense, highly turbulent medium. Within this medium, stellar feedback continually reorganises the gas, while gravitational collapse and star formation are triggered locally in the densest structures \citep{Wang2025-SFE}.
The resulting spectrum thus represents a regime in which star formation is tightly coupled to a shock-dominated, feedback-driven turbulent cascade. To illustrate this connection, Figure~\ref{fig:light} also shows the power spectrum of the cold ISM, defined by $T<10^2\K$, which approximately traces the turbulence-dominated gas component. We plot this spectrum with arbitrary amplitude because the reference fields in Equation~\eqref{eq:od} are not directly comparable. Its slope nevertheless agrees well with that of the H$\alpha$ spectrum.

By contrast, tracers sensitive to longer star formation timescales have progressively steeper spectra. This trend broadly agrees with the observations of nine local irregular galaxies by \citet{Willett2005}. On scales of $10$--$400\pc$, they found H$\alpha$ power-spectrum slopes spanning $\alpha\simeq-1.6$ to $\alpha\simeq0$, with a median of $\alpha\simeq-0.5$. The $V$-band images had systematically steeper slopes, ranging from $\alpha\simeq-1.3$ to $-1.5$, with a median of $\alpha\simeq-1.4$.

The dependence on wavelength arises because stars gradually lose the clustered spatial distribution inherited from their birth sites. Young stars remain concentrated within their natal star-forming regions, whereas older stellar populations have had more time to disperse through orbital motions and dynamical mixing, producing a progressively smoother spatial distribution. As a result, emission from older stellar populations contains less small-scale structure and exhibits a steeper power spectrum than tracers of ongoing star formation such as H$\alpha$. From a statistical perspective, this evolution suppresses small-scale intermittency and redistributes power towards larger scales. Consequently, the emission no longer traces individual star-forming clumps but instead reflects the cumulative stellar structure of the galaxy. The right panel of Figure~\ref{fig:light}, where the stellar mass is selected using cumulative age cuts, directly supports this interpretation. The spectra steepen progressively with stellar age, demonstrating how the dynamical evolution of stellar populations naturally smooths small-scale structure over time.

Overall, our results expose a potential discrepancy between the observed emission field and the stellar mass distribution. In this context, the stellar clumps recently revealed by \textit{JWST} in the rest-frame UV trace spatially concentrated recent star formation rather than long-lived stellar-mass substructures. 

\section{Discussion}\label{sec:discussions}
\subsection{Additional observational effects}\label{subsec:obs}

In real observations, additional instrumental and observational effects, such as the finite point spread function (PSF) and noise, can alter the apparent clumpiness. We assess their impact using the FUV mock images of the representative \thesanzoom galaxy ``m12.6'' at $z=6$ as a worked example (see the bottom-right panel of Figure~\ref{fig:vis}).

Following the approaches of e.g. \citet{LaChance2025,GuzmanOrtega2025}, we apply a series of post-processing steps to the FUV images. We first convert the intrinsic luminosity in each pixel to observed flux density using the luminosity distance, then map the physical pixel scale to angular units using the angular diameter distance at the corresponding redshift. We next convolve the images with the PSF appropriate for the observed-frame wavelength of rest-frame 1500~\AA\@. Over the redshift range considered here ($6\lesssim z\lesssim8$), this wavelength falls in the \textit{JWST}/NIRCam F115W filter. We generate the corresponding PSF models with \textsc{STPSF} \citep{Perrin2014}\footnote{\url{https://stpsf.readthedocs.io/en/latest/index.html}}.
We do not explicitly include the filter transmission curve, as we do not expect it to affect the inferred clumpiness significantly.

To account for noise, we first rebin the maps to a pixel scale of $0.03\,\mathrm{arcsec}$. We then add a constant background radiance of $0.25\,\mathrm{MJy\,sr^{-1}}$ and simulate photon noise by drawing uncorrelated Poisson realisations of the detected electron counts in each pixel. Because this realisation is drawn from the total signal rather than from the source alone, the resulting noise includes contributions from both the source and the background. The expected electron count in a given pixel is
\begin{equation}
    N_{\mathrm{pix,exp}} = \frac{I_{\mathrm{pix}}}{\mathrm{sens}_{\mathrm{avg}}}\times e_{\mathrm{gain}}\times t\times \mu\, ,
\end{equation}
where $\mathrm{sens}_{\mathrm{avg}}$ is the average pixel sensitivity in $(\mathrm{MJy\,sr^{-1}})/(\mathrm{DN\,s^{-1}})$, $I_{\mathrm{pix}}$ is the pixel radiance in $\mathrm{MJy\,sr^{-1}}$, $e_{\mathrm{gain}}$ is the conversion gain in $\mathrm{e^-\,DN^{-1}}$, and $t$ is the exposure time in seconds. We also introduce a magnification factor $\mu$ to mimic gravitational lensing. We subtract the expected background before computing the overdensity while retaining its Poisson fluctuations.

We adopt parameter values motivated by observations of Cosmic Grapes \citep{Fujimoto2025}. We take $t=5000\,\mathrm{s}$, $e_{\mathrm{gain}}=2.0$, and $\mu=32$. We set the average pixel sensitivity to $\mathrm{sens}_{\mathrm{avg}}=3.009\,(\mathrm{MJy\,sr^{-1}})/(\mathrm{DN\,s^{-1}})$, computed from the ``PHOTMJSR'' values in the \textit{JWST} calibration and zeropoint table (2023 September 14 release; \citealt{Gordon2022,Rieke2022,Rigby2023})\footnote{\url{https://jwst-docs.stsci.edu/files/216457093/216457096/1/1762453608965/NRC_ZPs_1126pmap.txt}}. We draw a single Poisson realisation of the electron count in each pixel and convert the resulting counts back to $\mathrm{MJy\,sr^{-1}}$.

\begin{figure}
    \centering
    \includegraphics[width=0.95\linewidth]{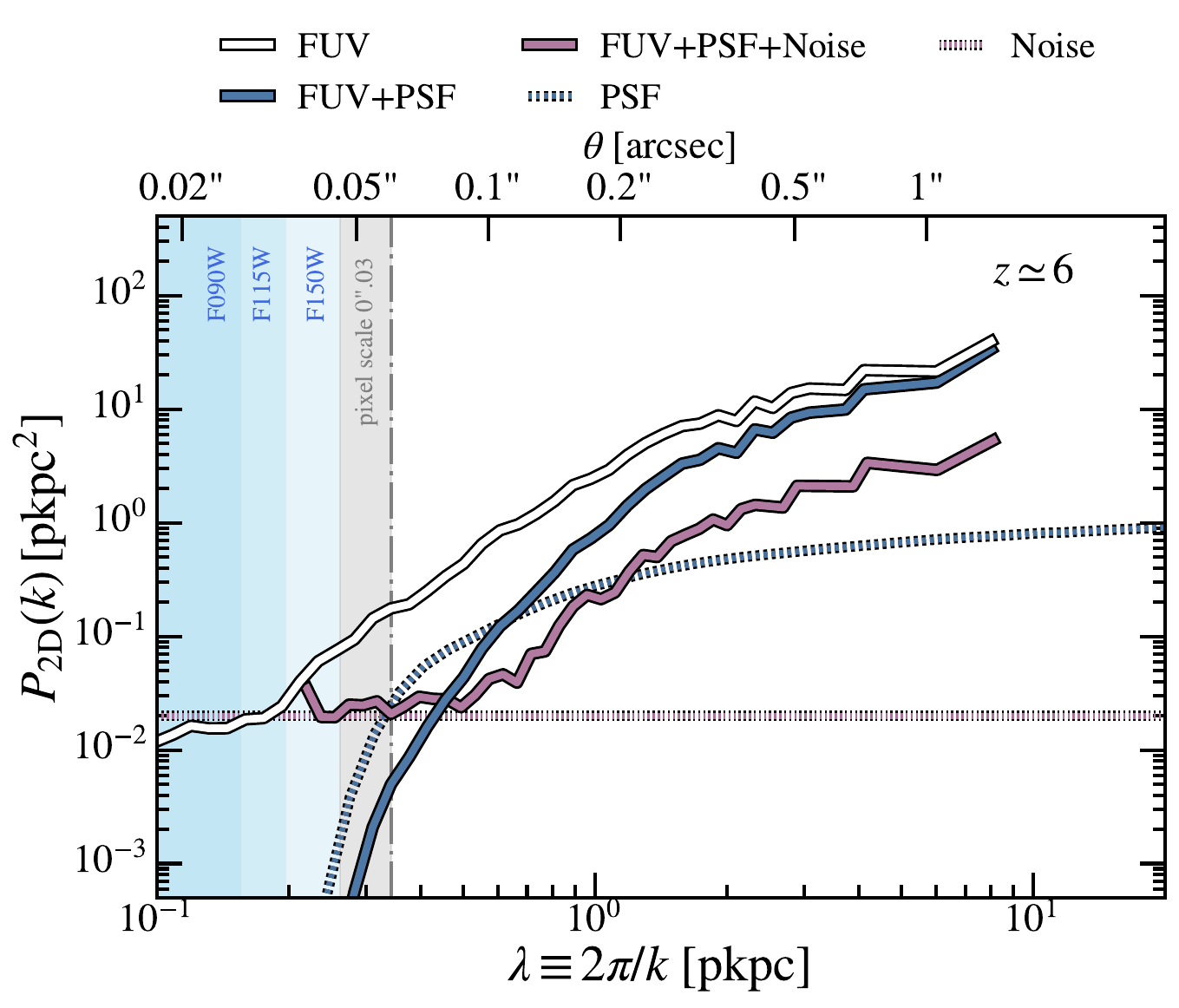}

    \caption{Rest-frame FUV power spectra before and after applying PSF convolution and noise for the representative \thesanzoom galaxy ``m12.6'' at $z=6$. We adopt the \textit{JWST}/NIRCam F150W PSF and the observational parameters described in Section~\ref{subsec:obs}. For reference, we show the F150W PSF transfer function and the power spectrum of pure Poisson noise. The coloured bands mark the PSF FWHM reference scales for the indicated filters, whereas the grey band marks the pixel Nyquist scale. The upper axis gives the corresponding angular scale at $z=6$.
    The PSF rapidly suppresses small-scale power, while noise introduces a white-noise component at the smallest resolved scales.}
    \label{fig:realism}
\end{figure}

Figure~\ref{fig:realism} presents the resulting spectra together with the PSF transfer function and the spectrum expected from pure Poisson noise. PSF convolution suppresses small-scale clumpiness over a broad range of scales. The signal begins to decline on scales larger than the PSF FWHM and becomes strongly attenuated towards smaller scales; the FWHM should therefore be treated as a reference scale rather than as a sharp resolution boundary.
The physical scale of this effect varies with redshift. Over the redshift range considered here, the angular-diameter distance decreases towards higher redshift, such that a fixed angular resolution corresponds to a smaller physical scale and the PSF-induced suppression shifts towards higher $k$.

In real space, PSF convolution broadens compact structures. For approximately Gaussian clumps, the observed characteristic radius $R_{\mathrm{obs}}$ is related to the intrinsic radius $R_{\mathrm{int}}$ by $R_{\mathrm{obs}}^2\simeq R_{\mathrm{int}}^2+R_{\mathrm{PSF}}^2$, where $R_{\mathrm{PSF}}$ characterises the PSF width. Marginally resolved clumps can consequently have upward-biased sizes, while neighbouring compact clumps separated by a distance comparable to the PSF scale can blend into a single larger complex \citep{Kalita2025}. In this case, the inferred clump abundance may be biased low, particularly near the effective resolution limit. More generally, the smallest clumps identified in resolved \textit{JWST} images should be interpreted as a PSF-filtered representation of the underlying emission field rather than as a direct census of intrinsic compact structures.

Poisson noise adds an approximately white noise-power term. On larger scales, it suppresses the signal through background dilution by an amount set by the signal-to-noise ratio.

\subsection{Impact of early stellar feedback}

We next examine how the power spectra depend on the subgrid physics implemented in the simulations, focusing in particular on the strength of stellar feedback. To isolate this dependence, we compare several \thesanzoom model variants with the independent \textsc{FIRE-2} high-redshift simulation suite \citep{Hopkins2018,Ma2018,Ma2019,Wetzel2023}\footnote{\url{http://flathub.flatironinstitute.org/fire}}.

The two simulation suites have broadly similar mass resolution and comparable prescriptions for star formation and stellar feedback, but differ in two main respects. First, the fiducial \thesanzoom run includes an empirical early stellar feedback (ESF) model, implemented by injecting momentum at $1000\,\mathrm{km\,s^{-1}\,Myr^{-1}}$ per unit stellar mass formed during the first $5\Myr$ after star formation \citep{Kannan2025}. This additional component is intended to improve agreement with the observed relation between stellar mass and halo mass at high redshift \citep[e.g.,][]{Tacchella2018,Behroozi2019}. It may also compensate effectively for physical processes not included in the simulations, such as cosmic rays \citep[e.g.,][]{Pakmor2016b,Buck2020,Hopkins2020-cr}, magnetic fields \citep[e.g.,][]{Marinacci2016,Hopkins2020-rad}, Lyman-$\alpha$ radiation pressure \citep[e.g.,][]{Smith2017,Kimm2018,Nebrin2025,Nebrin2026,Menon2026}, or other numerical uncertainties.
This component is removed in the corresponding noESF runs, and \textsc{FIRE-2} includes no analogous ESF prescription. Second, \textsc{FIRE-2} adopts a much smaller stellar gravitational softening length, $\epsilon_\ast=2.1\pc$, fixed in physical units. This value is chosen to match the minimum gravitational softening length of the gas cells, allowing newly formed stellar particles to inherit the small-scale structure of their parent gas clouds.

For the \textsc{FIRE-2} comparison, we select four galaxies, ``z5m12c'', ``z5m12a'', ``z5m11d'', and ``z5m11e'' from the high-$z$ runs \citep{Ma2018}. Their stellar masses are comparable to those of \thesanzoom galaxies at $z=6$, spanning $10^{7.5}$--$10^{9}\msun$. Because the \thesanzoom model variants are available only for lower-mass haloes, we instead select three main-target galaxies, ``m11.9'', ``m11.5'', and ``m11.1'', at $z=3$ thereby matching a similar stellar-mass range.

\begin{figure}
    \centering
    \includegraphics[width=0.95\linewidth]{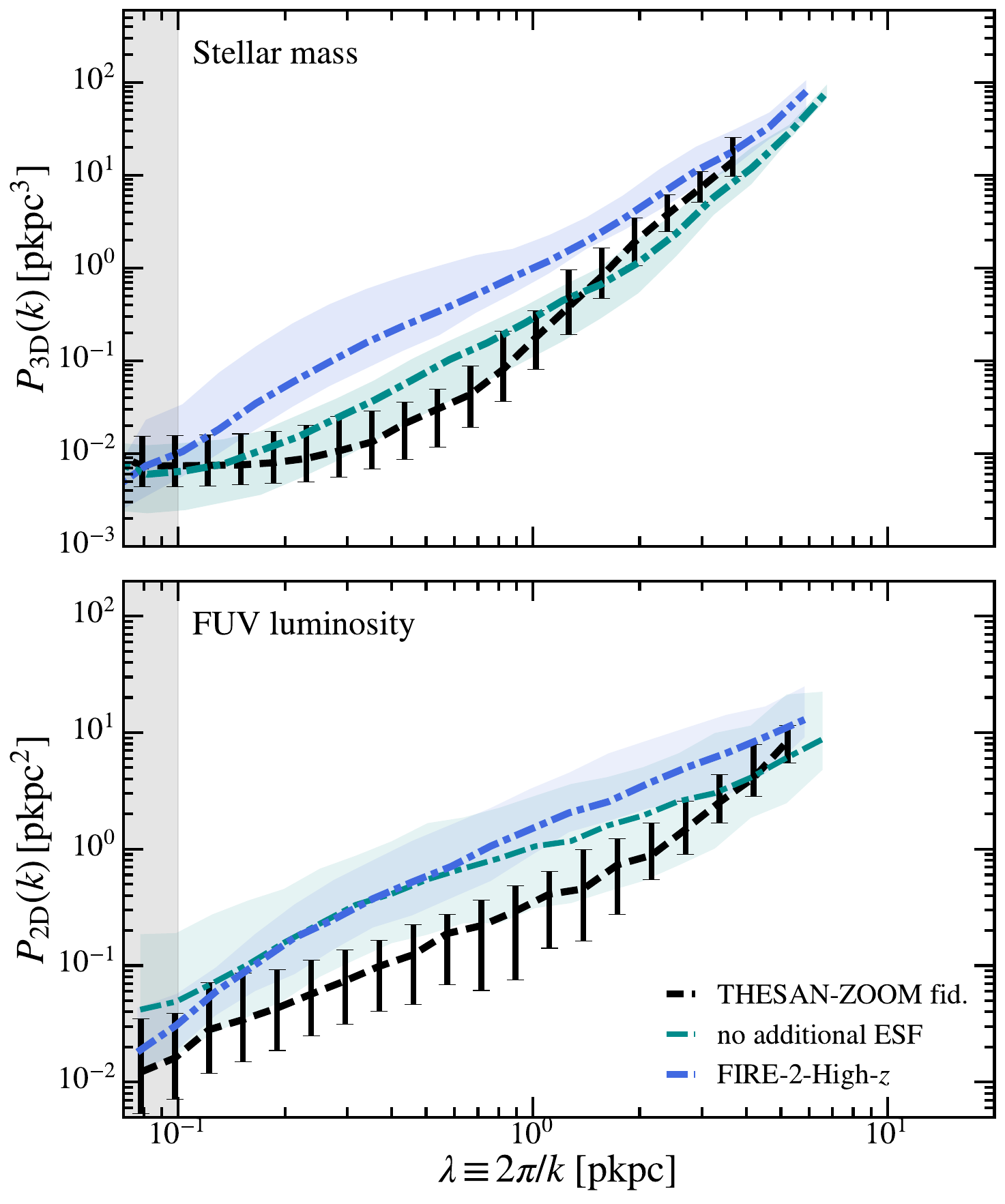}
    \caption{Mass (top) and FUV (bottom) power spectra for galaxies with comparable stellar masses in the fiducial \thesanzoom run, the noESF variant, and the high-redshift \textsc{FIRE-2} sample. Curves show the median spectra, and shaded regions and error bars denote the $16{\mathrm{th}}$--$84{\mathrm{th}}$ percentile variation. Without the additional ESF, small-scale power is enhanced, and the structure is more clumpy.}
    \label{fig:model}
\end{figure}

Figure~\ref{fig:model} shows the results. Removing ESF clearly enhances small-scale power and produces more pronounced clumpy structure. In the FUV, the noESF and \textsc{FIRE-2} models have remarkably similar spectra, whereas clear differences remain in the stellar mass distribution. The presence of early stellar feedback therefore accounts for a substantial part of the difference between the fiducial \thesanzoom and \textsc{FIRE-2} models, particularly in how they regulate the clumpiness of light-emitting structures. The residual differences probably arise from the much smaller stellar softening length in \textsc{FIRE-2}, as well as from other aspects of the respective numerical implementations.

Physically, this behaviour reflects how stellar feedback regulates the star-forming ISM and the subsequent evolution of newly formed stellar structures. First, feedback controls the efficiency with which dense gas becomes stars. \citet{Wang2025-SFE} showed that stronger early stellar feedback reduces the ambient gas density and lowers the star formation efficiency of individual giant molecular clouds while leaving the overall ISM cloud mass fraction largely unchanged. Individual star-forming regions are consequently less luminous, suppressing high-$k$ power in the emission maps. Second, after stars form, feedback continues to shape the spatial distribution of the young stellar population by dispersing gas, unbinding nascent clusters, and disrupting compact star-forming complexes before they can evolve into long-lived stellar structures. Together, these processes suppress small-scale power in both light and stellar mass. The reduced clumpiness of the fiducial \thesanzoom run therefore reflects efficient feedback acting both before star formation and during the subsequent evolution of young stellar structures.

\subsection{Connection to the bursty star formation history}
\begin{figure}
    \centering
    \includegraphics[width=0.95\linewidth]{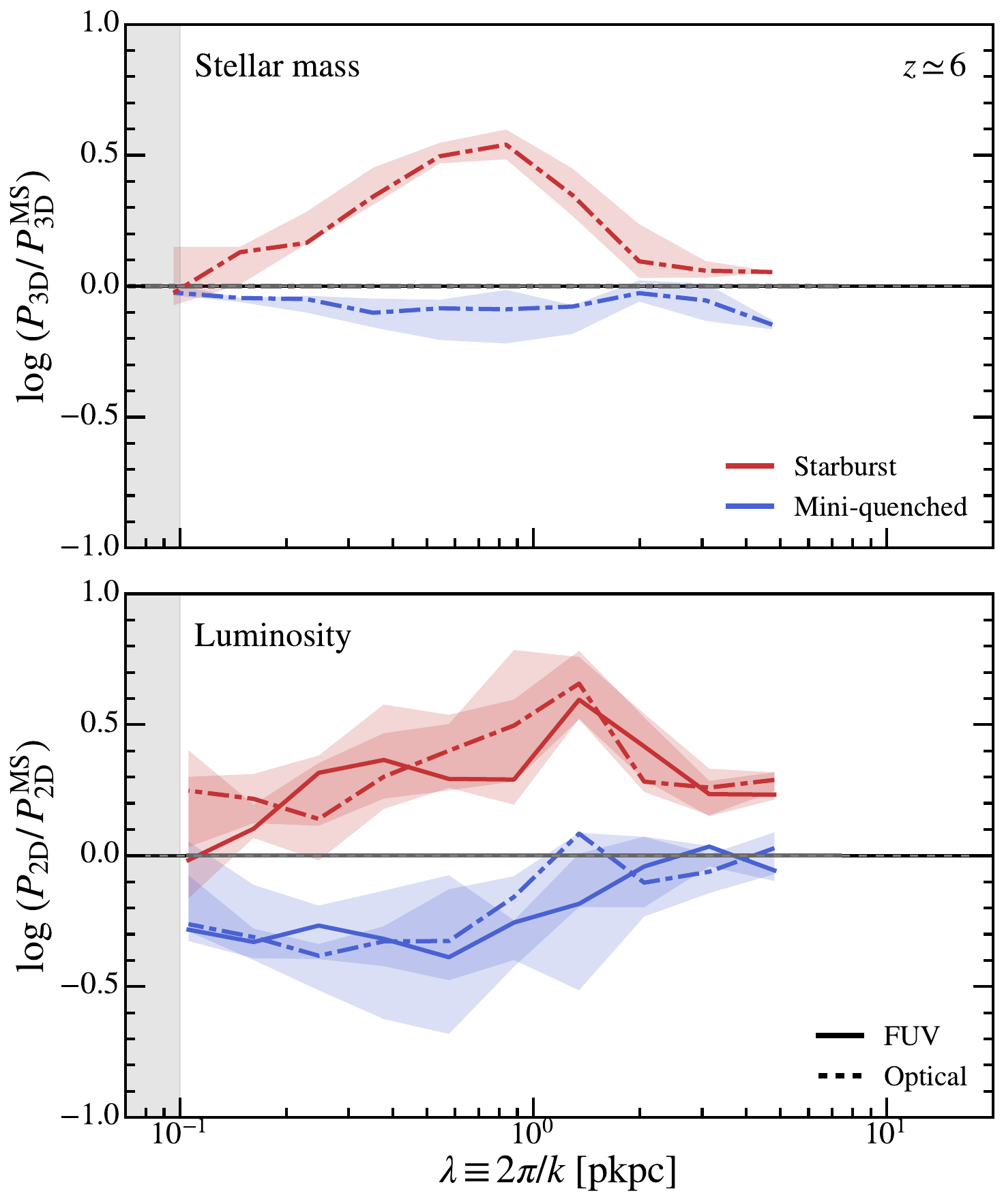}
    \caption{Power spectra relative to the main-sequence phase for galaxies at $z\simeq6$, shown for stellar mass (top) and emitted light (bottom). Shaded regions show the 16th–84th percentiles across galaxies. Curves show the starburst and mini-quenched phases, corresponding to the upper ($>84$th) and lower ($<16$th) percentiles of the sSFR distribution, respectively. The stellar mass spectra show enhanced small-scale power during starbursts, while the mini-quenched phase remains close to the main sequence. The light spectra vary more strongly across the star formation cycle, with enhanced small-scale power during starbursts and suppressed power during mini-quenching, indicating a closer connection between light clumpiness and recent star formation activity.}
    \label{fig:bursty}
\end{figure}

Although galaxy clumpiness depends sensitively on the stellar feedback model, its interpretation is complicated by the bursty nature of high-redshift star formation, as predicted by theoretical models \citep[e.g.,][]{Smit2016,Sparre2017,Tacchella2020,Shen2023,Sun2023b,Pallottini2023} and empirically found in observations \citep[e.g.,][]{Ciesla2024,Cole2025,Endsley2025,Munoz2026}. Because clumpy morphology is closely associated with recent star formation, galaxies observed at different phases of their star formation cycle may exhibit markedly different levels of clumpiness under the same physical model. To quantify this additional source of variation, we examine the temporal evolution of clumpiness over a typical duty cycle, focusing on starburst episodes around the target redshift.

We identify a starburst as a local peak in the SFR averaged over $30\Myr$ within one dynamical time, for which the peak SFR exceeds twice the median over that window. This definition is designed to capture a complete rise-and-decline cycle of star formation rather than only the peak itself. We then rank all snapshots within this interval by their sSFR and divide them into three phases according to the resulting percentile distribution. We define the central 16th–84th percentile as the main-sequence phase, the upper $>84$th percentile as the starburst phase, and the lower $<16$th percentile as the mini-quenched phase.

Figure~\ref{fig:bursty} presents the stacked stellar mass, FUV, and optical power spectra relative to the main-sequence phase, showing how clumpiness varies systematically across the burst cycle.
The mass spectrum is sensitive only to the strongest starburst phase, during which its small-scale power increases. The main-sequence and mini-quenched phases, by contrast, behave very similarly. The enhanced stellar mass clumpiness during starbursts is therefore relatively short-lived. Once the burst subsides, stellar clumps disperse or phase-mix on a timescale comparable to that required for the galaxy to return towards the main sequence, although the present analysis does not directly track individual clumps.
The FUV and optical emission respond more directly to the current star-forming phase. Their enhancement during starbursts is comparable to that in the stellar mass distribution, but their spectra show a much clearer suppression during mini-quenching.

One possible origin of the correlation between clumpiness and burstiness is that an upturn in star formation increases the fraction of newly formed stars that remain spatially concentrated in luminous stellar clumps. Based on an auxiliary examination of stellar clusters in \thesanzoom, we estimate characteristic stellar ages of $\sim10$–$50\Myr$ for individual clumps. This estimate is broadly consistent with the young ages inferred for observed clumps at $z>5$, which span a broader range extending to $\sim100\mathrm{Myr}$~\citep[e.g.,][]{Claeyssens2023}. The contribution of such clumps to the total stellar mass, and particularly to the rest-frame UV emission, may therefore approximately track the fraction of stars formed within the preceding few tens of Myr. This motivates a connection between clumpiness and the ratio of recent to longer-term star formation, $\mathrm{SFR}_{\tau}/\mathrm{SFR}_{100}$, where we adopt $\tau=30\Myr$ as an approximate timescale associated with the survival of young clumpy structures. High-redshift galaxies commonly exhibit strong SFR variability on sub-$100$-Myr time-scales \citep[e.g.,][]{Sparre2017,Looser2025}, allowing this ratio to reach $\sim2$--$3$ during a rising burst and thereby enhancing the prominence of stellar clumps. By contrast, normal spiral galaxies in the local Universe are closer to a quasi-steady regime on these time-scales, with ${\rm SFR}_{30}/{\rm SFR}_{100}\simeq1$, as motivated by the close agreement and relatively small scatter between their H$\alpha$- and FUV-based SFRs and by modelling of their recent SFHs \citep[e.g.,][]{Lee2009,Weisz2012,Karachentsev2018,Emami2019}. In our simulations, however, even the most massive galaxies remain in a relatively bursty star-forming regime down to $z=3$. Their persistently strong short-time-scale variability therefore sustains a high level of clumpiness and washes out an otherwise expected redshift evolution.

This result has important implications for the morphological bias of high-redshift galaxies. Several \textit{JWST}-based studies have measured the evolution of both the clumpy galaxy fraction and the fraction of light in clumps, confirming that clumpy systems remain common towards higher redshift \citep[e.g.,][]{Sattari2023,Vega2026}. Our results imply that selection effects could contribute to the observed trend. 
Because the intrinsic clumpiness of \thesanzoom galaxies evolves little with redshift, the observed increase in clumpy morphologies may be linked more closely to the star formation duty cycle than to redshift alone.
Observational incompleteness may therefore preferentially select galaxies caught in phases of elevated star formation, when their light distributions are both brighter and more clumpy.

\subsection{Comparison to observed clumpy galaxies}

While power-spectrum-based characterisations of galaxy clumpiness have not yet been widely applied to high-redshift observations, the clumpiness parameter $S$ is a commonly used alternative \citep{Conselice2003}. Whereas the power spectrum retains scale-dependent information, $S$ compresses the contribution of high-frequency structure into a single statistic, defined as
\begin{equation}
S = 10 \times
\frac{\sum_{\bm{j}}\left[\left(I_{\bm{j}}-I^{\sigma}_{\bm{j}}\right)-B_{\bm{j}}\right]}
{\sum_{\bm{j}}I_{\bm{j}}}\, ,
\label{eq:S}
\end{equation}
where $I_{\bm{j}}$ is the flux in pixel $\bm{j}=(x,y)$, $I^\sigma_{\bm{j}}$ is the corresponding image smoothed with a Gaussian filter of width $\sigma$, and $B_{\bm{j}}$ accounts for the background contribution estimated in the same manner. The sum includes all pixels assigned to the galaxy.
The $S$ statistic quantifies the fractional contribution of high-frequency structure to the total light distribution and can therefore be regarded as a compressed measure of small-scale power. In this section, we measure $S$ in our mock observations to enable a more direct comparison with existing observational studies that use this statistic \citep[e.g., Cosmic Grapes;][]{Fujimoto2025}.

We follow \citet{Fujimoto2025} to measure $S$ for the \thesanzoom galaxy ``m12.6'' over $5<z<8$. Its stellar mass in this redshift range, $M_\ast\simeq10^8$--$10^9\msun$, is comparable to that of the observed sample. We construct rest-frame FUV images from 10 randomly selected viewing angles within a $3\times3\kpc^2$ field of view, then convolve them with a \textit{JWST}/NIRCam F150W PSF to reproduce the instrumental response\footnote{We note that the rest-frame wavelengths are not exactly matched. At the redshift of the source studied by \citet{Fujimoto2025}, the F150W imaging probes $\simeq2140$\,{\AA}, whereas our mock images are constructed at $1500$\,{\AA}.}. We adopt an effective source-plane pixel scale of $0.01\,\mathrm{arcsec}$, corresponding to $\sim60\pc$ at $z\simeq6$, to match the source-plane spatial resolution achieved through gravitational lensing. We then add Poisson noise following Section~\ref{subsec:obs}. Finally, we measure $S$ using the definition and smoothing scale ($\sigma=200\pc$) adopted by \citet{Fujimoto2025}.

\begin{figure}
    \centering
    \includegraphics[width=1\linewidth]{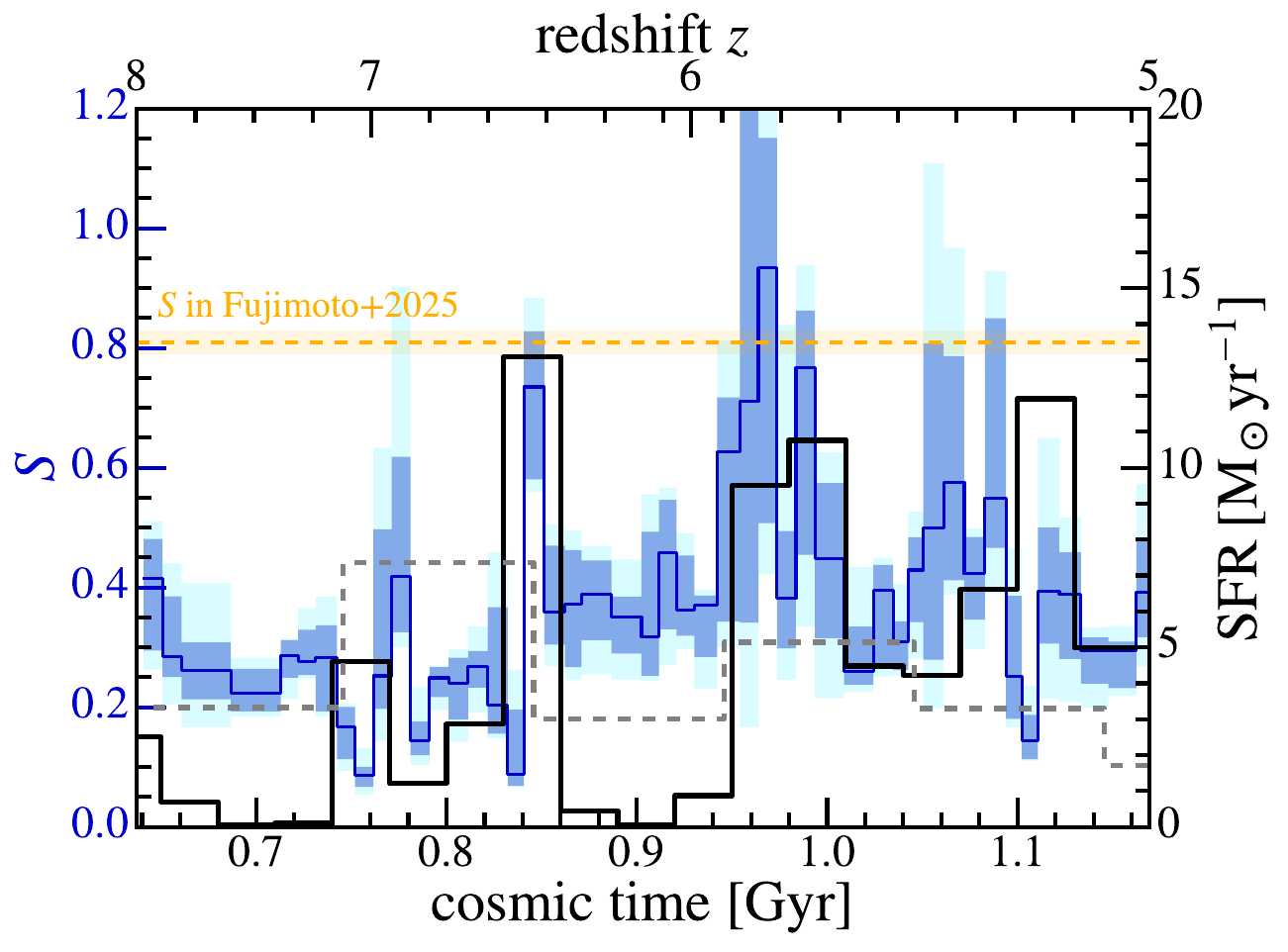}
    \caption{Evolution of the clumpiness parameter $S$ in the rest-frame FUV (blue; left axis) and the star formation history (black and gray; right axis) for the \thesanzoom galaxy ``m12.6’’ over the redshift range $5<z<8$. The solid blue curve shows the median $S$ measured from 10 randomly selected lines of sight, while the dark and light shaded regions indicate the $1\sigma$ and $2\sigma$ scatter, respectively. The horizontal dashed line and shaded band denote the observed value, $S=0.81\pm0.02$, measured for the Cosmic Grapes by \citet{Fujimoto2025}. The black solid and grey dashed curves show the SFR averaged over the preceding 30 Myr and 100 Myr, respectively. The clumpiness parameter exhibits strong temporal variability and closely follows the bursty star formation history, reaching values comparable to the observations only during the most intense starburst episodes.}
    \label{fig:S}
\end{figure}

Figure~\ref{fig:S} shows the temporal evolution of $S$ over $5<z<8$. The measured values exhibit substantial variability, fluctuating by nearly an order of magnitude over the star formation cycle. To facilitate comparison with recent star formation, we also show the SFR averaged over the preceding 30 Myr and 100 Myr. The former approximately matches the clump crossing timescale, while the latter matches the timescale over which \citet{Fujimoto2025} inferred the SFR from their observations. 

For this galaxy, the time series show an apparent temporal association between $S$ and the SFR. Periods of elevated SFR consistently coincide with enhanced clumpiness, indicating that galaxies become substantially more clumpy during starbursts. During the strongest bursts (with $\mathrm{sSFR}_{30}\gtrsim10\,\mathrm{Gyr}^{-1}$), $S$ reaches $\sim0.8$, consistent with the observational measurement of $S=0.81\pm0.02$ reported by \citet{Fujimoto2025}. At these times, the corresponding $\mathrm{sSFR}_{100}\simeq5\,\mathrm{Gyr}^{-1}$, similar to the value inferred for the Cosmic Grapes and well within the intrinsic scatter of the star-forming main sequence at these redshifts \citep{Iyer2018}. The clumpiness parameter is therefore highly time-dependent and closely tied to recent star formation, reinforcing the conclusion from the power-spectrum analysis that clumpy morphologies preferentially accompany active star-forming phases.

Reproducing the observed clumpiness does not, nonetheless, imply that \thesanzoom fully reproduces the underlying physical properties of the observed systems. ALMA kinematics indicate that the Cosmic Grapes hosts a rotationally supported gaseous disc with $V/\sigma_{\rm v}=3.58\pm0.74$ \citep{Fujimoto2025}, whereas we find no similarly rotation-dominated gas discs ($V/\sigma_{\rm v}>1$) in \thesanzoom at comparable redshifts. This discrepancy highlights the continuing challenge for modern cosmological simulations to simultaneously reproduce both the highly clumpy morphologies and the dynamically cold gaseous discs inferred for some high-redshift galaxies.

\section{Conclusions}

In this work, we have explored a new approach to quantifying the clumpiness of high-redshift galaxies in the \thesanzoom simulations. We analyse the auto-power spectra of both three-dimensional and projected stellar mass density fields, together with mock images of the FUV, optical, and H$\alpha$ emission. This comparison allows us to investigate the connection between the intrinsic morphologies of galaxies and their observable appearance. We further examine how the power spectra depend on star formation activity and on the underlying physical model. Our main findings are summarised below.

\begin{enumerate}
    \item The intrinsic stellar mass distributions of \thesanzoom galaxies remain relatively smooth across cosmic time, with no preferred Fourier mode. By contrast, the emission distributions remain highly structured and distinct from the smoother stellar mass distribution, showing substantially enhanced small-scale power. Clumps identified in observations therefore primarily trace the spatial distribution of recent star formation rather than the underlying stellar mass distribution.
    
    \item The light power spectra approximately follow power-law scalings from $P(k)\propto k^{-1}$ to $k^{-2}$, with shallower slopes for tracers sensitive to shorter star formation timescales. This behaviour is consistent with a highly compressible, shock-dominated medium, in which high-redshift star formation is concentrated in compact, short-lived, and strongly clustered sites shaped by feedback-driven turbulence.

    \item Observational effects further modify the measured clumpiness. PSF convolution strongly suppresses small-scale power, steepening the spectrum and producing a gradual suppression that becomes strong near and below the PSF scale. Poisson noise, on the other hand, adds a white-noise component near the pixel scale while reducing the signal on larger scales according to the signal-to-noise ratio.

    \item The strength of stellar feedback strongly influences the small-scale clustering power of both the stellar mass distribution and observable light. Comparing the fiducial and noESF \thesanzoom runs shows that reducing early stellar feedback enhances clumpiness and brings the light power spectra closer to those of \textsc{FIRE-2}. The remaining differences probably reflect numerical effects related to gravitational softening, including the different softening lengths adopted for newly formed stellar particles.
    
    \item Clumpiness varies across the star formation duty cycle. Stellar mass clumpiness increases during the most intense starburst phases, whereas light responds more strongly and immediately, with enhanced small-scale power during bursts and suppressed power during mini-quenching. \thesanzoom reproduces the high levels of clumpiness inferred from current observations during its strongest starburst phases. This suggests that observationally identified clumpy galaxies may preferentially represent systems caught during periods of elevated star formation activity.

\end{enumerate}

Overall, our results underscore the substantial gap between the intrinsic and apparent morphologies of high-redshift galaxies. Galaxy morphology remains a sensitive probe of galaxy formation physics, but its interpretation is complicated by the out-of-equilibrium state of galaxies across the star formation duty cycle and by the strong wavelength dependence of apparent structure. Because clumpiness varies substantially with star formation phase, apparent morphology should be used cautiously when constraining galaxy formation models. Future follow-up studies could directly trace the subsequent evolution of these clumpy structures and investigate their connection to dense star clusters and globular cluster progenitors using simulations capable of resolving and self-consistently modelling their internal dynamics.

\section*{Acknowledgements}
The authors gratefully acknowledge the Gauss Centre for Supercomputing e.V. (\url{www.gauss-centre.eu}) for funding this project by providing computing time on the GCS Supercomputer SuperMUC-NG at Leibniz Supercomputing Centre (\url{www.lrz.de}), under project pn29we.
The authors acknowledge the MIT Office of Research Computing and Data for providing resources that have contributed to the research results reported within this paper.
Support for programs JWST-AR-04814 (XS, MV) and JWST-AR-08709 (AS) was provided by NASA through a grant from the Space Telescope Science Institute, which is operated by the Association of Universities for Research in Astronomy, Inc., under NASA contract NAS 5-03127.
RK acknowledges support of the Natural Sciences and Engineering Research Council of Canada (NSERC) through a Discovery Grant and a Discovery Launch Supplement, funding reference numbers RGPIN-2024-06222 and DGECR-2024-00144.
EG is supported by the JSPS KAKENHI grant ILR 23K20035. 
LK acknowledges the support of a Royal Society University Research Fellowship (grant number URF$\backslash$R1$\backslash$251793).

\section*{Data Availability}
All simulation data, including snapshots, group catalogues, and merger trees, will be made publicly available in the near future via \url{www.thesan-project.com}. Before the public release, the data underlying this article can be shared upon reasonable request to the corresponding author(s).



\bibliographystyle{mnras}
\bibliography{reference} 

@ARTICLE{Kannan2025,
       author = {{Kannan}, Rahul and {Puchwein}, Ewald and {Smith}, Aaron and {Borrow}, Josh and {Garaldi}, Enrico and {Keating}, Laura and {Vogelsberger}, Mark and {Zier}, Oliver and {McClymont}, William and {Shen}, Xuejian and {Popovic}, Filip and {Tacchella}, Sandro and {Hernquist}, Lars and {Springel}, Volker},
        title = "{Introducing the THESAN-ZOOM project: radiation-hydrodynamic simulations of high-redshift galaxies with a multi-phase interstellar medium}",
      journal = {arXiv e-prints},
         year = 2025,
        month = feb,
          eid = {arXiv:2502.20437},
        pages = {arXiv:2502.20437},
archivePrefix = {arXiv},
       eprint = {2502.20437},
 primaryClass = {astro-ph.GA},
       adsurl = {https://ui.adsabs.harvard.edu/abs/2025arXiv250220437K}
}

@ARTICLE{Donnan2023,
       author = {{Donnan}, C.~T. and {McLeod}, D.~J. and {Dunlop}, J.~S. and {McLure}, R.~J. and {Carnall}, A.~C. and {Begley}, R. and {Cullen}, F. and {Hamadouche}, M.~L. and {Bowler}, R.~A.~A. and {Magee}, D. and {McCracken}, H.~J. and {Milvang-Jensen}, B. and {Moneti}, A. and {Targett}, T.},
        title = "{The evolution of the galaxy UV luminosity function at redshifts z ≃ 8 - 15 from deep JWST and ground-based near-infrared imaging}",
      journal = {\mnras},
         year = 2023,
        month = feb,
       volume = {518},
       number = {4},
        pages = {6011-6040},
          doi = {10.1093/mnras/stac3472},
archivePrefix = {arXiv},
       eprint = {2207.12356},
 primaryClass = {astro-ph.GA},
       adsurl = {https://ui.adsabs.harvard.edu/abs/2023MNRAS.518.6011D}
}

@ARTICLE{Ma2018,
       author = {{Ma}, Xiangcheng and {Hopkins}, Philip F. and {Garrison-Kimmel}, Shea and {Faucher-Gigu{\`e}re}, Claude-Andr{\'e} and {Quataert}, Eliot and {Boylan-Kolchin}, Michael and {Hayward}, Christopher C. and {Feldmann}, Robert and {Kere{\v{s}}}, Du{\v{s}}an},
        title = "{Simulating galaxies in the reionization era with FIRE-2: galaxy scaling relations, stellar mass functions, and luminosity functions}",
      journal = {\mnras},
         year = 2018,
        month = aug,
       volume = {478},
       number = {2},
        pages = {1694-1715},
          doi = {10.1093/mnras/sty1024},
archivePrefix = {arXiv},
       eprint = {1706.06605},
 primaryClass = {astro-ph.GA},
       adsurl = {https://ui.adsabs.harvard.edu/abs/2018MNRAS.478.1694M}
}

@ARTICLE{Pakmor2016b,
       author = {{Pakmor}, R. and {Pfrommer}, C. and {Simpson}, C.~M. and {Springel}, V.},
        title = "{Galactic Winds Driven by Isotropic and Anisotropic Cosmic-Ray Diffusion in Disk Galaxies}",
      journal = {\apjl},
         year = 2016,
        month = jun,
       volume = {824},
       number = {2},
          eid = {L30},
        pages = {L30},
          doi = {10.3847/2041-8205/824/2/L30},
archivePrefix = {arXiv},
       eprint = {1605.00643},
 primaryClass = {astro-ph.GA},
       adsurl = {https://ui.adsabs.harvard.edu/abs/2016ApJ...824L..30P}
}

@ARTICLE{Rizzo2020,
       author = {{Rizzo}, F. and {Vegetti}, S. and {Powell}, D. and {Fraternali}, F. and {McKean}, J.~P. and {Stacey}, H.~R. and {White}, S.~D.~M.},
        title = "{A dynamically cold disk galaxy in the early Universe}",
      journal = {\nat},
         year = 2020,
        month = aug,
       volume = {584},
       number = {7820},
        pages = {201-204},
          doi = {10.1038/s41586-020-2572-6},
archivePrefix = {arXiv},
       eprint = {2009.01251},
 primaryClass = {astro-ph.GA},
       adsurl = {https://ui.adsabs.harvard.edu/abs/2020Natur.584..201R}
}

@ARTICLE{RO2023,
       author = {{Roman-Oliveira}, Fernanda and {Fraternali}, Filippo and {Rizzo}, Francesca},
        title = "{Regular rotation and low turbulence in a diverse sample of z {\ensuremath{\sim}} 4.5 galaxies observed with ALMA}",
      journal = {\mnras},
         year = 2023,
        month = may,
       volume = {521},
       number = {1},
        pages = {1045-1065},
          doi = {10.1093/mnras/stad530},
archivePrefix = {arXiv},
       eprint = {2302.03049},
 primaryClass = {astro-ph.GA},
       adsurl = {https://ui.adsabs.harvard.edu/abs/2023MNRAS.521.1045R}
}

@ARTICLE{Rowland2024,
       author = {{Rowland}, Lucie E. and {Hodge}, Jacqueline and {Bouwens}, Rychard and {Pi{\~n}a}, Pavel E. Mancera and {Hygate}, Alexander and {Algera}, Hiddo and {Aravena}, Manuel and {Bowler}, Rebecca and {da Cunha}, Elisabete and {Dayal}, Pratika and {Ferrara}, Andrea and {Herard-Demanche}, Thomas and {Inami}, Hanae and {van Leeuwen}, Ivana and {de Looze}, Ilse and {Oesch}, Pascal and {Pallottini}, Andrea and {Phillips}, Si{\^a}n and {Rybak}, Matus and {Schouws}, Sander and {Smit}, Renske and {Sommovigo}, Laura and {Stefanon}, Mauro and {van der Werf}, Paul},
        title = "{REBELS-25: discovery of a dynamically cold disc galaxy at z = 7.31}",
      journal = {\mnras},
         year = 2024,
        month = dec,
       volume = {535},
       number = {3},
        pages = {2068-2091},
          doi = {10.1093/mnras/stae2217},
archivePrefix = {arXiv},
       eprint = {2405.06025},
 primaryClass = {astro-ph.GA},
       adsurl = {https://ui.adsabs.harvard.edu/abs/2024MNRAS.535.2068R}
}

@ARTICLE{Mager2018,
       author = {{Mager}, Violet A. and {Conselice}, Christopher J. and {Seibert}, Mark and {Gusbar}, Courtney and {Katona}, Anthony P. and {Villari}, Joseph M. and {Madore}, Barry F. and {Windhorst}, Rogier A.},
        title = "{Galaxy Structure in the Ultraviolet: The Dependence of Morphological Parameters on Rest-frame Wavelength}",
      journal = {\apj},
         year = 2018,
        month = sep,
       volume = {864},
       number = {2},
          eid = {123},
        pages = {123},
          doi = {10.3847/1538-4357/aad59e},
archivePrefix = {arXiv},
       eprint = {1808.00577},
 primaryClass = {astro-ph.GA},
       adsurl = {https://ui.adsabs.harvard.edu/abs/2018ApJ...864..123M}
}

@article{Ma2019,
   title={Dust attenuation, dust emission, and dust temperature in galaxies at z ≥ 5: a view from the FIRE-2 simulations},
   volume={487},
   ISSN={1365-2966},
   url={http://dx.doi.org/10.1093/mnras/stz1324},
   DOI={10.1093/mnras/stz1324},
   number={2},
   journal={Monthly Notices of the Royal Astronomical Society},
   publisher={Oxford University Press (OUP)},
   author={Ma, Xiangcheng and Hayward, Christopher C and Casey, Caitlin M and Hopkins, Philip F and Quataert, Eliot and Liang, Lichen and Faucher-Giguère, Claude-André and Feldmann, Robert and Kereš, Dušan},
   year={2019},
   month=May, pages={1844–1864} }

@ARTICLE{Shin2023,
       author = {{Shin}, Eun-jin and {Tacchella}, Sandro and {Kim}, Ji-hoon and {Iyer}, Kartheik G. and {Semenov}, Vadim A.},
        title = "{Star Formation Variability as a Probe for the Baryon Cycle within Galaxies}",
      journal = {\apj},
         year = 2023,
        month = apr,
       volume = {947},
       number = {2},
          eid = {61},
        pages = {61},
          doi = {10.3847/1538-4357/acc251},
archivePrefix = {arXiv},
       eprint = {2211.01922},
 primaryClass = {astro-ph.GA},
       adsurl = {https://ui.adsabs.harvard.edu/abs/2023ApJ...947...61S}
}

@ARTICLE{Mandelker2017,
       author = {{Mandelker}, Nir and {Dekel}, Avishai and {Ceverino}, Daniel and {DeGraf}, Colin and {Guo}, Yicheng and {Primack}, Joel},
        title = "{Giant clumps in simulated high- z Galaxies: properties, evolution and dependence on feedback}",
      journal = {\mnras},
         year = 2017,
        month = jan,
       volume = {464},
       number = {1},
        pages = {635-665},
          doi = {10.1093/mnras/stw2358},
archivePrefix = {arXiv},
       eprint = {1512.08791},
 primaryClass = {astro-ph.GA},
       adsurl = {https://ui.adsabs.harvard.edu/abs/2017MNRAS.464..635M}
}

@ARTICLE{Willett2005,
       author = {{Willett}, Kyle W. and {Elmegreen}, Bruce G. and {Hunter}, Deidre A.},
        title = "{Power Spectra in V band and H{\ensuremath{\alpha}} of Nine Irregular Galaxies}",
      journal = {\aj},
         year = 2005,
        month = may,
       volume = {129},
       number = {5},
        pages = {2186-2196},
          doi = {10.1086/429678},
archivePrefix = {arXiv},
       eprint = {astro-ph/0503295},
 primaryClass = {astro-ph},
       adsurl = {https://ui.adsabs.harvard.edu/abs/2005AJ....129.2186W}
}

@ARTICLE{Madau2014,
       author = {{Madau}, Piero and {Weisz}, Daniel R. and {Conroy}, Charlie},
        title = "{Reversal of Fortune: Increased Star Formation Efficiencies in the Early Histories of Dwarf Galaxies?}",
      journal = {\apjl},
         year = 2014,
        month = aug,
       volume = {790},
       number = {2},
          eid = {L17},
        pages = {L17},
          doi = {10.1088/2041-8205/790/2/L17},
archivePrefix = {arXiv},
       eprint = {1406.0838},
 primaryClass = {astro-ph.GA},
       adsurl = {https://ui.adsabs.harvard.edu/abs/2014ApJ...790L..17M}
}

@ARTICLE{Springel2021,
       author = {{Springel}, Volker and {Pakmor}, R{\"u}diger and {Zier}, Oliver and {Reinecke}, Martin},
        title = "{Simulating cosmic structure formation with the GADGET-4 code}",
      journal = {\mnras},
         year = 2021,
        month = sep,
       volume = {506},
       number = {2},
        pages = {2871-2949},
          doi = {10.1093/mnras/stab1855},
archivePrefix = {arXiv},
       eprint = {2010.03567},
 primaryClass = {astro-ph.IM},
       adsurl = {https://ui.adsabs.harvard.edu/abs/2021MNRAS.506.2871S}
}

@ARTICLE{Tacchella2020,
       author = {{Tacchella}, Sandro and {Forbes}, John C. and {Caplar}, Neven},
        title = "{Stochastic modelling of star-formation histories II: star-formation variability from molecular clouds and gas inflow}",
      journal = {\mnras},
         year = 2020,
        month = sep,
       volume = {497},
       number = {1},
        pages = {698-725},
          doi = {10.1093/mnras/staa1838},
archivePrefix = {arXiv},
       eprint = {2006.09382},
 primaryClass = {astro-ph.GA},
       adsurl = {https://ui.adsabs.harvard.edu/abs/2020MNRAS.497..698T}
}

@incollection{Burgers1948,
title = {A Mathematical Model Illustrating the Theory of Turbulence},
editor = {Richard {Von Mises} and Theodore {Von Kármán}},
series = {Advances in Applied Mechanics},
publisher = {Elsevier},
volume = {1},
pages = {171-199},
year = {1948},
issn = {0065-2156},
doi = {https://doi.org/10.1016/S0065-2156(08)70100-5},
url = {https://www.sciencedirect.com/science/article/pii/S0065215608701005},
author = {J.M. Burgers}
}

@ARTICLE{Dekel2009,
       author = {{Dekel}, Avishai and {Sari}, Re'em and {Ceverino}, Daniel},
        title = "{Formation of Massive Galaxies at High Redshift: Cold Streams, Clumpy Disks, and Compact Spheroids}",
      journal = {\apj},
         year = 2009,
        month = sep,
       volume = {703},
       number = {1},
        pages = {785-801},
          doi = {10.1088/0004-637X/703/1/785},
archivePrefix = {arXiv},
       eprint = {0901.2458},
 primaryClass = {astro-ph.GA},
       adsurl = {https://ui.adsabs.harvard.edu/abs/2009ApJ...703..785D}
}

@ARTICLE{Forster2011,
       author = {{F{\"o}rster Schreiber}, N.~M. and {Shapley}, A.~E. and {Erb}, D.~K. and {Genzel}, R. and {Steidel}, C.~C. and {Bouch{\'e}}, N. and {Cresci}, G. and {Davies}, R.},
        title = "{Constraints on the Assembly and Dynamics of Galaxies. I. Detailed Rest-frame Optical Morphologies on Kiloparsec Scale of z \raisebox{-0.5ex}\textasciitilde 2 Star-forming Galaxies}",
      journal = {\apj},
         year = 2011,
        month = apr,
       volume = {731},
       number = {1},
          eid = {65},
        pages = {65},
          doi = {10.1088/0004-637X/731/1/65},
archivePrefix = {arXiv},
       eprint = {1011.1507},
 primaryClass = {astro-ph.CO},
       adsurl = {https://ui.adsabs.harvard.edu/abs/2011ApJ...731...65F}
}

@ARTICLE{Elmegreen2009,
       author = {{Elmegreen}, Bruce G. and {Elmegreen}, Debra Meloy and {Fernandez}, Maria Ximena and {Lemonias}, Jenna Jo},
        title = "{Bulge and Clump Evolution in Hubble Ultra Deep Field Clump Clusters, Chains and Spiral Galaxies}",
      journal = {\apj},
         year = 2009,
        month = feb,
       volume = {692},
       number = {1},
        pages = {12-31},
          doi = {10.1088/0004-637X/692/1/12},
archivePrefix = {arXiv},
       eprint = {0810.5404},
 primaryClass = {astro-ph},
       adsurl = {https://ui.adsabs.harvard.edu/abs/2009ApJ...692...12E}
}

@ARTICLE{Bournaud2007,
       author = {{Bournaud}, Fr{\'e}d{\'e}ric and {Elmegreen}, Bruce G. and {Elmegreen}, Debra Meloy},
        title = "{Rapid Formation of Exponential Disks and Bulges at High Redshift from the Dynamical Evolution of Clump-Cluster and Chain Galaxies}",
      journal = {\apj},
         year = 2007,
        month = nov,
       volume = {670},
       number = {1},
        pages = {237-248},
          doi = {10.1086/522077},
archivePrefix = {arXiv},
       eprint = {0708.0306},
 primaryClass = {astro-ph},
       adsurl = {https://ui.adsabs.harvard.edu/abs/2007ApJ...670..237B}
}

@ARTICLE{Sun2023b,
       author = {{Sun}, Guochao and {Faucher-Gigu{\`e}re}, Claude-Andr{\'e} and {Hayward}, Christopher C. and {Shen}, Xuejian and {Wetzel}, Andrew and {Cochrane}, Rachel K.},
        title = "{Bursty Star Formation Naturally Explains the Abundance of Bright Galaxies at Cosmic Dawn}",
      journal = {\apjl},
         year = 2023,
        month = oct,
       volume = {955},
       number = {2},
          eid = {L35},
        pages = {L35},
          doi = {10.3847/2041-8213/acf85a},
archivePrefix = {arXiv},
       eprint = {2307.15305},
 primaryClass = {astro-ph.GA},
       adsurl = {https://ui.adsabs.harvard.edu/abs/2023ApJ...955L..35S}
}

@ARTICLE{Hopkins2020-rad,
       author = {{Hopkins}, Philip F. and {Grudi{\'c}}, Michael Y. and {Wetzel}, Andrew and {Kere{\v{s}}}, Du{\v{s}}an and {Faucher-Gigu{\`e}re}, Claude-Andr{\'e} and {Ma}, Xiangcheng and {Murray}, Norman and {Butcher}, Nathan},
        title = "{Radiative stellar feedback in galaxy formation: Methods and physics}",
      journal = {\mnras},
         year = 2020,
        month = jan,
       volume = {491},
       number = {3},
        pages = {3702-3729},
          doi = {10.1093/mnras/stz3129},
archivePrefix = {arXiv},
       eprint = {1811.12462},
 primaryClass = {astro-ph.GA},
       adsurl = {https://ui.adsabs.harvard.edu/abs/2020MNRAS.491.3702H}
}

@ARTICLE{Marinacci2016,
       author = {{Marinacci}, Federico and {Vogelsberger}, Mark},
        title = "{Effects of simulated cosmological magnetic fields on the galaxy population}",
      journal = {\mnras},
         year = 2016,
        month = feb,
       volume = {456},
       number = {1},
        pages = {L69-L73},
          doi = {10.1093/mnrasl/slv176},
archivePrefix = {arXiv},
       eprint = {1508.06631},
 primaryClass = {astro-ph.CO},
       adsurl = {https://ui.adsabs.harvard.edu/abs/2016MNRAS.456L..69M}
}

@ARTICLE{Nebrin2025,
       author = {{Nebrin}, Olof and {Smith}, Aaron and {Lorinc}, Kevin and {H{\"o}rnquist}, Johan and {Larson}, {\r{A}}sa and {Mellema}, Garrelt and {Giri}, Sambit K.},
        title = "{Lyman-{\ensuremath{\alpha}} feedback prevails at Cosmic Dawn: Implications for the first galaxies, stars, and star clusters}",
      journal = {\mnras},
         year = 2025,
        month = jan,
          doi = {10.1093/mnras/staf038},
archivePrefix = {arXiv},
       eprint = {2409.19288},
 primaryClass = {astro-ph.GA},
       adsurl = {https://ui.adsabs.harvard.edu/abs/2025MNRAS.tmp...39N}
}

@ARTICLE{Nebrin2026,
       author = {{Nebrin}, Olof and {Smith}, Aaron and {Mellema}, Garrelt and {Lorinc}, Kevin and {Manzoni}, Daniele},
        title = "{Lyman-alpha Pressure Strongly Enhances Pre-Supernova Feedback at Cosmic Dawn: The First Multi-Dimensional Lyman-alpha Radiation Hydrodynamics Simulations}",
      journal = {arXiv e-prints},
         year = 2026,
        month = jun,
          eid = {arXiv:2606.02711},
        pages = {arXiv:2606.02711},
          doi = {10.48550/arXiv.2606.02711},
archivePrefix = {arXiv},
       eprint = {2606.02711},
 primaryClass = {astro-ph.GA},
       adsurl = {https://ui.adsabs.harvard.edu/abs/2026arXiv260602711N}
}

@ARTICLE{Menon2026,
       author = {{Menon}, Shyam H and {Smith}, Aaron},
        title = "{Lyman-alpha Radiation Pressure in Dense Star Clusters: Implications for Star Formation and Winds at Cosmic Dawn}",
      journal = {arXiv e-prints},
         year = 2026,
        month = may,
          eid = {arXiv:2605.13982},
        pages = {arXiv:2605.13982},
          doi = {10.48550/arXiv.2605.13982},
archivePrefix = {arXiv},
       eprint = {2605.13982},
 primaryClass = {astro-ph.GA},
       adsurl = {https://ui.adsabs.harvard.edu/abs/2026arXiv260513982M}
}

@ARTICLE{Kimm2018,
       author = {{Kimm}, Taysun and {Haehnelt}, Martin and {Blaizot}, J{\'e}r{\'e}my and {Katz}, Harley and {Michel-Dansac}, L{\'e}o and {Garel}, Thibault and {Rosdahl}, Joakim and {Teyssier}, Romain},
        title = "{Impact of Lyman alpha pressure on metal-poor dwarf galaxies}",
      journal = {\mnras},
         year = 2018,
        month = apr,
       volume = {475},
       number = {4},
        pages = {4617-4635},
          doi = {10.1093/mnras/sty126},
archivePrefix = {arXiv},
       eprint = {1801.04952},
 primaryClass = {astro-ph.GA},
       adsurl = {https://ui.adsabs.harvard.edu/abs/2018MNRAS.475.4617K}
}

@ARTICLE{Smith2017,
       author = {{Smith}, Aaron and {Bromm}, Volker and {Loeb}, Abraham},
        title = "{Lyman {\ensuremath{\alpha}} radiation hydrodynamics of galactic winds before cosmic reionization}",
      journal = {\mnras},
         year = 2017,
        month = jan,
       volume = {464},
       number = {3},
        pages = {2963-2978},
          doi = {10.1093/mnras/stw2591},
archivePrefix = {arXiv},
       eprint = {1607.07166},
 primaryClass = {astro-ph.GA},
       adsurl = {https://ui.adsabs.harvard.edu/abs/2017MNRAS.464.2963S}
}

@ARTICLE{Buck2020,
       author = {{Buck}, Tobias and {Pfrommer}, Christoph and {Pakmor}, R{\"u}diger and {Grand}, Robert J.~J. and {Springel}, Volker},
        title = "{The effects of cosmic rays on the formation of Milky Way-mass galaxies in a cosmological context}",
      journal = {\mnras},
         year = 2020,
        month = sep,
       volume = {497},
       number = {2},
        pages = {1712-1737},
          doi = {10.1093/mnras/staa1960},
archivePrefix = {arXiv},
       eprint = {1911.00019},
 primaryClass = {astro-ph.GA},
       adsurl = {https://ui.adsabs.harvard.edu/abs/2020MNRAS.497.1712B}
}

@ARTICLE{Hopkins2020-cr,
       author = {{Hopkins}, Philip F. and {Chan}, T.~K. and {Garrison-Kimmel}, Shea and {Ji}, Suoqing and {Su}, Kung-Yi and {Hummels}, Cameron B. and {Kere{\v{s}}}, Du{\v{s}}an and {Quataert}, Eliot and {Faucher-Gigu{\`e}re}, Claude-Andr{\'e}},
        title = "{But what about...: cosmic rays, magnetic fields, conduction, and viscosity in galaxy formation}",
      journal = {\mnras},
         year = 2020,
        month = mar,
       volume = {492},
       number = {3},
        pages = {3465-3498},
          doi = {10.1093/mnras/stz3321},
archivePrefix = {arXiv},
       eprint = {1905.04321},
 primaryClass = {astro-ph.GA},
       adsurl = {https://ui.adsabs.harvard.edu/abs/2020MNRAS.492.3465H}
}

@ARTICLE{Marinacci2019,
       author = {{Marinacci}, Federico and {Sales}, Laura V. and {Vogelsberger}, Mark and {Torrey}, Paul and {Springel}, Volker},
        title = "{Simulating the interstellar medium and stellar feedback on a moving mesh: implementation and isolated galaxies}",
      journal = {\mnras},
         year = 2019,
        month = nov,
       volume = {489},
       number = {3},
        pages = {4233-4260},
          doi = {10.1093/mnras/stz2391},
archivePrefix = {arXiv},
       eprint = {1905.08806},
 primaryClass = {astro-ph.GA},
       adsurl = {https://ui.adsabs.harvard.edu/abs/2019MNRAS.489.4233M}
}

@ARTICLE{Pallottini2023,
       author = {{Pallottini}, A. and {Ferrara}, A.},
        title = "{Stochastic star formation in early galaxies: Implications for the James Webb Space Telescope}",
      journal = {\aap},
         year = 2023,
        month = sep,
       volume = {677},
          eid = {L4},
        pages = {L4},
          doi = {10.1051/0004-6361/202347384},
archivePrefix = {arXiv},
       eprint = {2307.03219},
 primaryClass = {astro-ph.GA},
       adsurl = {https://ui.adsabs.harvard.edu/abs/2023A&A...677L...4P}
}

@ARTICLE{Hopkins2018,
       author = {{Hopkins}, Philip F. and {Wetzel}, Andrew and {Kere{\v{s}}}, Du{\v{s}}an and {Faucher-Gigu{\`e}re}, Claude-Andr{\'e} and {Quataert}, Eliot and {Boylan-Kolchin}, Michael and {Murray}, Norman and {Hayward}, Christopher C. and {Garrison-Kimmel}, Shea and {Hummels}, Cameron and {Feldmann}, Robert and {Torrey}, Paul and {Ma}, Xiangcheng and {Angl{\'e}s-Alc{\'a}zar}, Daniel and {Su}, Kung-Yi and {Orr}, Matthew and {Schmitz}, Denise and {Escala}, Ivanna and {Sanderson}, Robyn and {Grudi{\'c}}, Michael Y. and {Hafen}, Zachary and {Kim}, Ji-Hoon and {Fitts}, Alex and {Bullock}, James S. and {Wheeler}, Coral and {Chan}, T.~K. and {Elbert}, Oliver D. and {Narayanan}, Desika},
        title = "{FIRE-2 simulations: physics versus numerics in galaxy formation}",
      journal = {\mnras},
         year = 2018,
        month = oct,
       volume = {480},
       number = {1},
        pages = {800-863},
          doi = {10.1093/mnras/sty1690},
archivePrefix = {arXiv},
       eprint = {1702.06148},
 primaryClass = {astro-ph.GA},
       adsurl = {https://ui.adsabs.harvard.edu/abs/2018MNRAS.480..800H}
}

@ARTICLE{Hopkins2014,
       author = {{Hopkins}, Philip F. and {Kere{\v{s}}}, Du{\v{s}}an and {O{\~n}orbe}, Jos{\'e} and {Faucher-Gigu{\`e}re}, Claude-Andr{\'e} and {Quataert}, Eliot and {Murray}, Norman and {Bullock}, James S.},
        title = "{Galaxies on FIRE (Feedback In Realistic Environments): stellar feedback explains cosmologically inefficient star formation}",
      journal = {\mnras},
         year = 2014,
        month = nov,
       volume = {445},
       number = {1},
        pages = {581-603},
          doi = {10.1093/mnras/stu1738},
archivePrefix = {arXiv},
       eprint = {1311.2073},
 primaryClass = {astro-ph.CO},
       adsurl = {https://ui.adsabs.harvard.edu/abs/2014MNRAS.445..581H}
}

@ARTICLE{Zier2024,
       author = {{Zier}, Oliver and {Kannan}, Rahul and {Smith}, Aaron and {Vogelsberger}, Mark and {Verbeek}, Erkin},
        title = "{Adapting AREPO-RT for exascale computing: GPU acceleration and efficient communication}",
      journal = {\mnras},
         year = 2024,
        month = sep,
       volume = {533},
       number = {1},
        pages = {268-286},
          doi = {10.1093/mnras/stae1837},
archivePrefix = {arXiv},
       eprint = {2404.17630},
 primaryClass = {astro-ph.IM},
       adsurl = {https://ui.adsabs.harvard.edu/abs/2024MNRAS.533..268Z}
}

@ARTICLE{Tacchella2018,
       author = {{Tacchella}, Sandro and {Bose}, Sownak and {Conroy}, Charlie and {Eisenstein}, Daniel J. and {Johnson}, Benjamin D.},
        title = "{A Redshift-independent Efficiency Model: Star Formation and Stellar Masses in Dark Matter Halos at z {\ensuremath{\gtrsim}} 4}",
      journal = {\apj},
         year = 2018,
        month = dec,
       volume = {868},
       number = {2},
          eid = {92},
        pages = {92},
          doi = {10.3847/1538-4357/aae8e0},
archivePrefix = {arXiv},
       eprint = {1806.03299},
 primaryClass = {astro-ph.GA},
       adsurl = {https://ui.adsabs.harvard.edu/abs/2018ApJ...868...92T}
}

@ARTICLE{Shen2024c-size,
       author = {{Shen}, Xuejian and {Vogelsberger}, Mark and {Borrow}, Josh and {Hu}, Yongao and {Erickson}, Evan and {Kannan}, Rahul and {Smith}, Aaron and {Garaldi}, Enrico and {Hernquist}, Lars and {Morishita}, Takahiro and {Tacchella}, Sandro and {Zier}, Oliver and {Sun}, Guochao and {Eilers}, Anna-Christina and {Wang}, Hui},
        title = "{The THESAN project: galaxy sizes during the epoch of reionization}",
      journal = {\mnras},
         year = 2024,
        month = oct,
       volume = {534},
       number = {2},
        pages = {1433-1458},
          doi = {10.1093/mnras/stae2156},
archivePrefix = {arXiv},
       eprint = {2402.08717},
 primaryClass = {astro-ph.GA},
       adsurl = {https://ui.adsabs.harvard.edu/abs/2024MNRAS.534.1433S}
}

@ARTICLE{Shen2024b-ede,
       author = {{Shen}, Xuejian and {Vogelsberger}, Mark and {Boylan-Kolchin}, Michael and {Tacchella}, Sandro and {Naidu}, Rohan P.},
        title = "{Early galaxies and early dark energy: a unified solution to the hubble tension and puzzles of massive bright galaxies revealed by JWST}",
      journal = {\mnras},
         year = 2024,
        month = oct,
       volume = {533},
       number = {4},
        pages = {3923-3936},
          doi = {10.1093/mnras/stae1932},
archivePrefix = {arXiv},
       eprint = {2406.15548},
 primaryClass = {astro-ph.GA},
       adsurl = {https://ui.adsabs.harvard.edu/abs/2024MNRAS.533.3923S}
}

@ARTICLE{Shen2023,
       author = {{Shen}, Xuejian and {Vogelsberger}, Mark and {Boylan-Kolchin}, Michael and {Tacchella}, Sandro and {Kannan}, Rahul},
        title = "{The impact of UV variability on the abundance of bright galaxies at z {\ensuremath{\geq}} 9}",
      journal = {\mnras},
         year = 2023,
        month = nov,
       volume = {525},
       number = {3},
        pages = {3254-3261},
          doi = {10.1093/mnras/stad2508},
archivePrefix = {arXiv},
       eprint = {2305.05679},
 primaryClass = {astro-ph.GA},
       adsurl = {https://ui.adsabs.harvard.edu/abs/2023MNRAS.525.3254S}
}

@ARTICLE{Davis1985,
       author = {{Davis}, M. and {Efstathiou}, G. and {Frenk}, C.~S. and {White}, S.~D.~M.},
        title = "{The evolution of large-scale structure in a universe dominated by cold dark matter}",
      journal = {\apj},
         year = 1985,
        month = may,
       volume = {292},
        pages = {371-394},
          doi = {10.1086/163168},
       adsurl = {https://ui.adsabs.harvard.edu/abs/1985ApJ...292..371D}
}

@ARTICLE{Springel2003a,
       author = {{Springel}, Volker and {Hernquist}, Lars},
        title = "{Cosmological smoothed particle hydrodynamics simulations: a hybrid multiphase model for star formation}",
      journal = {\mnras},
         year = 2003,
        month = feb,
       volume = {339},
       number = {2},
        pages = {289-311},
          doi = {10.1046/j.1365-8711.2003.06206.x},
archivePrefix = {arXiv},
       eprint = {astro-ph/0206393},
 primaryClass = {astro-ph},
       adsurl = {https://ui.adsabs.harvard.edu/abs/2003MNRAS.339..289S}
}

@ARTICLE{Planck2016,
       author = {{Planck Collaboration} and {Ade}, P.~A.~R. and {Aghanim}, N. and {Arnaud}, M. and {Ashdown}, M. and {Aumont}, J. and {Baccigalupi}, C. and {Banday}, A.~J. and {Barreiro}, R.~B. and {Bartlett}, J.~G. and {Bartolo}, N. and {Battaner}, E. and {Battye}, R. and {Benabed}, K. and {Beno{\^\i}t}, A. and {Benoit-L{\'e}vy}, A. and {Bernard}, J. -P. and {Bersanelli}, M. and {Bielewicz}, P. and {Bock}, J.~J. and {Bonaldi}, A. and {Bonavera}, L. and {Bond}, J.~R. and {Borrill}, J. and {Bouchet}, F.~R. and {Boulanger}, F. and {Bucher}, M. and {Burigana}, C. and {Butler}, R.~C. and {Calabrese}, E. and {Cardoso}, J. -F. and {Catalano}, A. and {Challinor}, A. and {Chamballu}, A. and {Chary}, R. -R. and {Chiang}, H.~C. and {Chluba}, J. and {Christensen}, P.~R. and {Church}, S. and {Clements}, D.~L. and {Colombi}, S. and {Colombo}, L.~P.~L. and {Combet}, C. and {Coulais}, A. and {Crill}, B.~P. and {Curto}, A. and {Cuttaia}, F. and {Danese}, L. and {Davies}, R.~D. and {Davis}, R.~J. and {de Bernardis}, P. and {de Rosa}, A. and {de Zotti}, G. and {Delabrouille}, J. and {D{\'e}sert}, F. -X. and {Di Valentino}, E. and {Dickinson}, C. and {Diego}, J.~M. and {Dolag}, K. and {Dole}, H. and {Donzelli}, S. and {Dor{\'e}}, O. and {Douspis}, M. and {Ducout}, A. and {Dunkley}, J. and {Dupac}, X. and {Efstathiou}, G. and {Elsner}, F. and {En{\ss}lin}, T.~A. and {Eriksen}, H.~K. and {Farhang}, M. and {Fergusson}, J. and {Finelli}, F. and {Forni}, O. and {Frailis}, M. and {Fraisse}, A.~A. and {Franceschi}, E. and {Frejsel}, A. and {Galeotta}, S. and {Galli}, S. and {Ganga}, K. and {Gauthier}, C. and {Gerbino}, M. and {Ghosh}, T. and {Giard}, M. and {Giraud-H{\'e}raud}, Y. and {Giusarma}, E. and {Gjerl{\o}w}, E. and {Gonz{\'a}lez-Nuevo}, J. and {G{\'o}rski}, K.~M. and {Gratton}, S. and {Gregorio}, A. and {Gruppuso}, A. and {Gudmundsson}, J.~E. and {Hamann}, J. and {Hansen}, F.~K. and {Hanson}, D. and {Harrison}, D.~L. and {Helou}, G. and {Henrot-Versill{\'e}}, S. and {Hern{\'a}ndez-Monteagudo}, C. and {Herranz}, D. and {Hildebrandt}, S.~R. and {Hivon}, E. and {Hobson}, M. and {Holmes}, W.~A. and {Hornstrup}, A. and {Hovest}, W. and {Huang}, Z. and {Huffenberger}, K.~M. and {Hurier}, G. and {Jaffe}, A.~H. and {Jaffe}, T.~R. and {Jones}, W.~C. and {Juvela}, M. and {Keih{\"a}nen}, E. and {Keskitalo}, R. and {Kisner}, T.~S. and {Kneissl}, R. and {Knoche}, J. and {Knox}, L. and {Kunz}, M. and {Kurki-Suonio}, H. and {Lagache}, G. and {L{\"a}hteenm{\"a}ki}, A. and {Lamarre}, J. -M. and {Lasenby}, A. and {Lattanzi}, M. and {Lawrence}, C.~R. and {Leahy}, J.~P. and {Leonardi}, R. and {Lesgourgues}, J. and {Levrier}, F. and {Lewis}, A. and {Liguori}, M. and {Lilje}, P.~B. and {Linden-V{\o}rnle}, M. and {L{\'o}pez-Caniego}, M. and {Lubin}, P.~M. and {Mac{\'\i}as-P{\'e}rez}, J.~F. and {Maggio}, G. and {Maino}, D. and {Mandolesi}, N. and {Mangilli}, A. and {Marchini}, A. and {Maris}, M. and {Martin}, P.~G. and {Martinelli}, M. and {Mart{\'\i}nez-Gonz{\'a}lez}, E. and {Masi}, S. and {Matarrese}, S. and {McGehee}, P. and {Meinhold}, P.~R. and {Melchiorri}, A. and {Melin}, J. -B. and {Mendes}, L. and {Mennella}, A. and {Migliaccio}, M. and {Millea}, M. and {Mitra}, S. and {Miville-Desch{\^e}nes}, M. -A. and {Moneti}, A. and {Montier}, L. and {Morgante}, G. and {Mortlock}, D. and {Moss}, A. and {Munshi}, D. and {Murphy}, J.~A. and {Naselsky}, P. and {Nati}, F. and {Natoli}, P. and {Netterfield}, C.~B. and {N{\o}rgaard-Nielsen}, H.~U. and {Noviello}, F. and {Novikov}, D. and {Novikov}, I. and {Oxborrow}, C.~A. and {Paci}, F. and {Pagano}, L. and {Pajot}, F. and {Paladini}, R. and {Paoletti}, D. and {Partridge}, B. and {Pasian}, F. and {Patanchon}, G. and {Pearson}, T.~J. and {Perdereau}, O. and {Perotto}, L. and {Perrotta}, F. and {Pettorino}, V. and {Piacentini}, F. and {Piat}, M. and {Pierpaoli}, E. and {Pietrobon}, D. and {Plaszczynski}, S. and {Pointecouteau}, E. and {Polenta}, G. and {Popa}, L. and {Pratt}, G.~W. and {Pr{\'e}zeau}, G. and {Prunet}, S. and {Puget}, J. -L. and {Rachen}, J.~P. and {Reach}, W.~T. and {Rebolo}, R. and {Reinecke}, M. and {Remazeilles}, M. and {Renault}, C. and {Renzi}, A. and {Ristorcelli}, I. and {Rocha}, G. and {Rosset}, C. and {Rossetti}, M. and {Roudier}, G. and {Rouill{\'e} d'Orfeuil}, B. and {Rowan-Robinson}, M. and {Rubi{\~n}o-Mart{\'\i}n}, J.~A. and {Rusholme}, B. and {Said}, N. and {Salvatelli}, V. and {Salvati}, L. and {Sandri}, M. and {Santos}, D. and {Savelainen}, M. and {Savini}, G. and {Scott}, D. and {Seiffert}, M.~D. and {Serra}, P. and {Shellard}, E.~P.~S. and {Spencer}, L.~D. and {Spinelli}, M. and {Stolyarov}, V. and {Stompor}, R. and {Sudiwala}, R. and {Sunyaev}, R. and {Sutton}, D. and {Suur-Uski}, A. -S. and {Sygnet}, J. -F. and {Tauber}, J.~A. and {Terenzi}, L. and {Toffolatti}, L. and {Tomasi}, M. and {Tristram}, M. and {Trombetti}, T. and {Tucci}, M. and {Tuovinen}, J. and {T{\"u}rler}, M. and {Umana}, G. and {Valenziano}, L. and {Valiviita}, J. and {Van Tent}, F. and {Vielva}, P. and {Villa}, F. and {Wade}, L.~A. and {Wandelt}, B.~D. and {Wehus}, I.~K. and {White}, M. and {White}, S.~D.~M. and {Wilkinson}, A. and {Yvon}, D. and {Zacchei}, A. and {Zonca}, A.},
        title = "{Planck 2015 results. XIII. Cosmological parameters}",
      journal = {\aap},
         year = 2016,
        month = sep,
       volume = {594},
          eid = {A13},
        pages = {A13},
          doi = {10.1051/0004-6361/201525830},
archivePrefix = {arXiv},
       eprint = {1502.01589},
 primaryClass = {astro-ph.CO},
       adsurl = {https://ui.adsabs.harvard.edu/abs/2016A&A...594A..13P}
}

@ARTICLE{Smith2022,
       author = {{Smith}, A. and {Kannan}, R. and {Garaldi}, E. and {Vogelsberger}, M. and {Pakmor}, R. and {Springel}, V. and {Hernquist}, L.},
        title = "{The THESAN project: Lyman-{\ensuremath{\alpha}} emission and transmission during the Epoch of Reionization}",
      journal = {\mnras},
         year = 2022,
        month = may,
       volume = {512},
       number = {3},
        pages = {3243-3265},
          doi = {10.1093/mnras/stac713},
archivePrefix = {arXiv},
       eprint = {2110.02966},
 primaryClass = {astro-ph.CO},
       adsurl = {https://ui.adsabs.harvard.edu/abs/2022MNRAS.512.3243S}
}

@ARTICLE{Garaldi2024,
       author = {{Garaldi}, Enrico and {Kannan}, Rahul and {Smith}, Aaron and {Borrow}, Josh and {Vogelsberger}, Mark and {Pakmor}, R{\"u}diger and {Springel}, Volker and {Hernquist}, Lars and {Gal{\'a}rraga-Espinosa}, Daniela and {Yeh}, Jessica Y. -C. and {Shen}, Xuejian and {Xu}, Clara and {Neyer}, Meredith and {Spina}, Benedetta and {Almualla}, Mouza and {Zhao}, Yu},
        title = "{The THESAN project: public data release of radiation-hydrodynamic simulations matching reionization-era JWST observations}",
      journal = {\mnras},
         year = 2024,
        month = jun,
       volume = {530},
       number = {4},
        pages = {3765-3786},
          doi = {10.1093/mnras/stae839},
archivePrefix = {arXiv},
       eprint = {2309.06475},
 primaryClass = {astro-ph.CO},
       adsurl = {https://ui.adsabs.harvard.edu/abs/2024MNRAS.530.3765G}
}

@ARTICLE{Garaldi2022,
       author = {{Garaldi}, E. and {Kannan}, R. and {Smith}, A. and {Springel}, V. and {Pakmor}, R. and {Vogelsberger}, M. and {Hernquist}, L.},
        title = "{The THESAN project: properties of the intergalactic medium and its connection to reionization-era galaxies}",
      journal = {\mnras},
         year = 2022,
        month = jun,
       volume = {512},
       number = {4},
        pages = {4909-4933},
          doi = {10.1093/mnras/stac257},
archivePrefix = {arXiv},
       eprint = {2110.01628},
 primaryClass = {astro-ph.CO},
       adsurl = {https://ui.adsabs.harvard.edu/abs/2022MNRAS.512.4909G}
}

@ARTICLE{Eldridge2017,
       author = {{Eldridge}, J.~J. and {Stanway}, E.~R. and {Xiao}, L. and {McClelland}, L.~A.~S. and {Taylor}, G. and {Ng}, M. and {Greis}, S.~M.~L. and {Bray}, J.~C.},
        title = "{Binary Population and Spectral Synthesis Version 2.1: Construction, Observational Verification, and New Results}",
      journal = {\pasa},
         year = 2017,
        month = nov,
       volume = {34},
          eid = {e058},
        pages = {e058},
          doi = {10.1017/pasa.2017.51},
archivePrefix = {arXiv},
       eprint = {1710.02154},
 primaryClass = {astro-ph.SR},
       adsurl = {https://ui.adsabs.harvard.edu/abs/2017PASA...34...58E}
}

@ARTICLE{Chabrier2003,
       author = {{Chabrier}, Gilles},
        title = "{Galactic Stellar and Substellar Initial Mass Function}",
      journal = {\pasp},
         year = 2003,
        month = jul,
       volume = {115},
       number = {809},
        pages = {763-795},
          doi = {10.1086/376392},
archivePrefix = {arXiv},
       eprint = {astro-ph/0304382},
 primaryClass = {astro-ph},
       adsurl = {https://ui.adsabs.harvard.edu/abs/2003PASP..115..763C}
}

@ARTICLE{Kannan2020,
       author = {{Kannan}, Rahul and {Marinacci}, Federico and {Vogelsberger}, Mark and {Sales}, Laura V. and {Torrey}, Paul and {Springel}, Volker and {Hernquist}, Lars},
        title = "{Simulating the interstellar medium of galaxies with radiative transfer, non-equilibrium thermochemistry, and dust}",
      journal = {\mnras},
         year = 2020,
        month = dec,
       volume = {499},
       number = {4},
        pages = {5732-5748},
          doi = {10.1093/mnras/staa3249},
archivePrefix = {arXiv},
       eprint = {1910.14041},
 primaryClass = {astro-ph.GA},
       adsurl = {https://ui.adsabs.harvard.edu/abs/2020MNRAS.499.5732K}
}

@ARTICLE{Kannan2019,
       author = {{Kannan}, Rahul and {Vogelsberger}, Mark and {Marinacci}, Federico and {McKinnon}, Ryan and {Pakmor}, R{\"u}diger and {Springel}, Volker},
        title = "{AREPO-RT: radiation hydrodynamics on a moving mesh}",
      journal = {\mnras},
         year = 2019,
        month = may,
       volume = {485},
       number = {1},
        pages = {117-149},
          doi = {10.1093/mnras/stz287},
archivePrefix = {arXiv},
       eprint = {1804.01987},
 primaryClass = {astro-ph.IM},
       adsurl = {https://ui.adsabs.harvard.edu/abs/2019MNRAS.485..117K}
}

@ARTICLE{Kannan2022thesan,
       author = {{Kannan}, R. and {Garaldi}, E. and {Smith}, A. and {Pakmor}, R. and {Springel}, V. and {Vogelsberger}, M. and {Hernquist}, L.},
        title = "{Introducing the THESAN project: radiation-magnetohydrodynamic simulations of the epoch of reionization}",
      journal = {\mnras},
         year = 2022,
        month = apr,
       volume = {511},
       number = {3},
        pages = {4005-4030},
          doi = {10.1093/mnras/stab3710},
archivePrefix = {arXiv},
       eprint = {2110.00584},
 primaryClass = {astro-ph.GA},
       adsurl = {https://ui.adsabs.harvard.edu/abs/2022MNRAS.511.4005K}
}

@ARTICLE{Springel2010,
       author = {{Springel}, Volker},
        title = "{E pur si muove: Galilean-invariant cosmological hydrodynamical simulations on a moving mesh}",
      journal = {\mnras},
         year = 2010,
        month = jan,
       volume = {401},
       number = {2},
        pages = {791-851},
          doi = {10.1111/j.1365-2966.2009.15715.x},
archivePrefix = {arXiv},
       eprint = {0901.4107},
 primaryClass = {astro-ph.CO},
       adsurl = {https://ui.adsabs.harvard.edu/abs/2010MNRAS.401..791S}
}

@ARTICLE{Behroozi2019,
       author = {{Behroozi}, Peter and {Wechsler}, Risa H. and {Hearin}, Andrew P. and {Conroy}, Charlie},
        title = "{UNIVERSEMACHINE: The correlation between galaxy growth and dark matter halo assembly from z = 0-10}",
      journal = {\mnras},
         year = 2019,
        month = sep,
       volume = {488},
       number = {3},
        pages = {3143-3194},
          doi = {10.1093/mnras/stz1182},
archivePrefix = {arXiv},
       eprint = {1806.07893},
 primaryClass = {astro-ph.GA},
       adsurl = {https://ui.adsabs.harvard.edu/abs/2019MNRAS.488.3143B}
}

@ARTICLE{Harikane2023,
       author = {{Harikane}, Yuichi and {Ouchi}, Masami and {Oguri}, Masamune and {Ono}, Yoshiaki and {Nakajima}, Kimihiko and {Isobe}, Yuki and {Umeda}, Hiroya and {Mawatari}, Ken and {Zhang}, Yechi},
        title = "{A Comprehensive Study of Galaxies at z   9-16 Found in the Early JWST Data: Ultraviolet Luminosity Functions and Cosmic Star Formation History at the Pre-reionization Epoch}",
      journal = {\apjs},
         year = 2023,
        month = mar,
       volume = {265},
       number = {1},
          eid = {5},
        pages = {5},
          doi = {10.3847/1538-4365/acaaa9},
archivePrefix = {arXiv},
       eprint = {2208.01612},
 primaryClass = {astro-ph.GA},
       adsurl = {https://ui.adsabs.harvard.edu/abs/2023ApJS..265....5H}
}

@ARTICLE{Cowie1995,
       author = {{Cowie}, Lennox L. and {Hu}, Esther M. and {Songaila}, Antoinette},
        title = "{Faintest Galaxy Morphologies From HST WFPC2 Imaging of the Hawaii Survey Fields}",
      journal = {\aj},
         year = 1995,
        month = oct,
       volume = {110},
        pages = {1576},
          doi = {10.1086/117631},
archivePrefix = {arXiv},
       eprint = {astro-ph/9507055},
 primaryClass = {astro-ph},
       adsurl = {https://ui.adsabs.harvard.edu/abs/1995AJ....110.1576C}
}

@ARTICLE{Giavalisco1996,
       author = {{Giavalisco}, Mauro and {Steidel}, Charles C. and {Macchetto}, F. Duccio},
        title = "{Hubble Space Telescope Imaging of Star-forming Galaxies at Redshifts Z > 3}",
      journal = {\apj},
         year = 1996,
        month = oct,
       volume = {470},
        pages = {189},
          doi = {10.1086/177859},
archivePrefix = {arXiv},
       eprint = {astro-ph/9603062},
 primaryClass = {astro-ph},
       adsurl = {https://ui.adsabs.harvard.edu/abs/1996ApJ...470..189G}
}

@ARTICLE{vandenBergh1996,
       author = {{van den Bergh}, Sidney},
        title = "{The Extragalactic Distance Scale}",
      journal = {\pasp},
         year = 1996,
        month = dec,
       volume = {108},
        pages = {1091-1096},
          doi = {10.1086/133839},
archivePrefix = {arXiv},
       eprint = {astro-ph/9604070},
 primaryClass = {astro-ph},
       adsurl = {https://ui.adsabs.harvard.edu/abs/1996PASP..108.1091V}
}

@INPROCEEDINGS{Elmegreen2005,
       author = {{Elmegreen}, D.~M. and {Elmegreen}, B.~G. and {Kaufman}, M. and {Sheth}, K. and {Struck}, C. and {Thomasson}, M. and {Brinks}, E.},
        title = "{Spitzer Space Telescope IRAC and MIPS Observations of the Interacting Galaxies IC2163 and NGC2207: Clumpy Emission}",
    booktitle = {American Astronomical Society Meeting Abstracts},
         year = 2005,
       series = {American Astronomical Society Meeting Abstracts},
       volume = {207},
        month = dec,
          eid = {128.14},
        pages = {128.14},
       adsurl = {https://ui.adsabs.harvard.edu/abs/2005AAS...20712814E}
}

@INPROCEEDINGS{Elmegreen2007,
       author = {{Elmegreen}, Debra Meloy},
        title = "{Photometric Properties of Clumpy Galaxies in the Hubble Ultra Deep Field}",
    booktitle = {Island Universes},
         year = 2007,
       editor = {{DE JONG}, R.~S.},
       series = {Astrophysics and Space Science Proceedings},
       volume = {3},
        month = jan,
        pages = {527},
          doi = {10.1007/978-1-4020-5573-7_91},
       adsurl = {https://ui.adsabs.harvard.edu/abs/2007ASSP....3..527E}
}

@ARTICLE{Elmegreen2008,
       author = {{Elmegreen}, Bruce G. and {Bournaud}, Fr{\'e}d{\'e}ric and {Elmegreen}, Debra Meloy},
        title = "{Nuclear Black Hole Formation in Clumpy Galaxies at High Redshift}",
      journal = {\apj},
         year = 2008,
        month = sep,
       volume = {684},
       number = {2},
        pages = {829-834},
          doi = {10.1086/590361},
archivePrefix = {arXiv},
       eprint = {0805.2266},
 primaryClass = {astro-ph},
       adsurl = {https://ui.adsabs.harvard.edu/abs/2008ApJ...684..829E}
}

@ARTICLE{Guo2012,
       author = {{Guo}, Yicheng and {Giavalisco}, Mauro and {Ferguson}, Henry C. and {Cassata}, Paolo and {Koekemoer}, Anton M.},
        title = "{Multi-wavelength View of Kiloparsec-scale Clumps in Star-forming Galaxies at z \raisebox{-0.5ex}\textasciitilde 2}",
      journal = {\apj},
         year = 2012,
        month = oct,
       volume = {757},
       number = {2},
          eid = {120},
        pages = {120},
          doi = {10.1088/0004-637X/757/2/120},
archivePrefix = {arXiv},
       eprint = {1110.3800},
 primaryClass = {astro-ph.CO},
       adsurl = {https://ui.adsabs.harvard.edu/abs/2012ApJ...757..120G}
}

@ARTICLE{Guo2015,
       author = {{Guo}, Yicheng and {Ferguson}, Henry C. and {Bell}, Eric F. and {Koo}, David C. and {Conselice}, Christopher J. and {Giavalisco}, Mauro and {Kassin}, Susan and {Lu}, Yu and {Lucas}, Ray and {Mandelker}, Nir and {McIntosh}, Daniel H. and {Primack}, Joel R. and {Ravindranath}, Swara and {Barro}, Guillermo and {Ceverino}, Daniel and {Dekel}, Avishai and {Faber}, Sandra M. and {Fang}, Jerome J. and {Koekemoer}, Anton M. and {Noeske}, Kai and {Rafelski}, Marc and {Straughn}, Amber},
        title = "{Clumpy Galaxies in CANDELS. I. The Definition of UV Clumps and the Fraction of Clumpy Galaxies at 0.5 < z < 3}",
      journal = {\apj},
         year = 2015,
        month = feb,
       volume = {800},
       number = {1},
          eid = {39},
        pages = {39},
          doi = {10.1088/0004-637X/800/1/39},
archivePrefix = {arXiv},
       eprint = {1410.7398},
 primaryClass = {astro-ph.GA},
       adsurl = {https://ui.adsabs.harvard.edu/abs/2015ApJ...800...39G}
}

@ARTICLE{Wisnioski2012,
       author = {{Wisnioski}, Emily and {Glazebrook}, Karl and {Blake}, Chris and {Poole}, Gregory B. and {Green}, Andrew W. and {Wyder}, Ted and {Martin}, Chris},
        title = "{Scaling relations of star-forming regions: from kpc-sized clumps to H II regions}",
      journal = {\mnras},
         year = 2012,
        month = jun,
       volume = {422},
       number = {4},
        pages = {3339-3355},
          doi = {10.1111/j.1365-2966.2012.20850.x},
archivePrefix = {arXiv},
       eprint = {1203.0309},
 primaryClass = {astro-ph.CO},
       adsurl = {https://ui.adsabs.harvard.edu/abs/2012MNRAS.422.3339W}
}

@ARTICLE{Wuyts2012,
       author = {{Wuyts}, Stijn and {F{\"o}rster Schreiber}, Natascha M. and {Genzel}, Reinhard and {Guo}, Yicheng and {Barro}, Guillermo and {Bell}, Eric F. and {Dekel}, Avishai and {Faber}, Sandra M. and {Ferguson}, Henry C. and {Giavalisco}, Mauro and {Grogin}, Norman A. and {Hathi}, Nimish P. and {Huang}, Kuang-Han and {Kocevski}, Dale D. and {Koekemoer}, Anton M. and {Koo}, David C. and {Lotz}, Jennifer and {Lutz}, Dieter and {McGrath}, Elizabeth and {Newman}, Jeffrey A. and {Rosario}, David and {Saintonge}, Amelie and {Tacconi}, Linda J. and {Weiner}, Benjamin J. and {van der Wel}, Arjen},
        title = "{Smooth(er) Stellar Mass Maps in CANDELS: Constraints on the Longevity of Clumps in High-redshift Star-forming Galaxies}",
      journal = {\apj},
         year = 2012,
        month = jul,
       volume = {753},
       number = {2},
          eid = {114},
        pages = {114},
          doi = {10.1088/0004-637X/753/2/114},
archivePrefix = {arXiv},
       eprint = {1203.2611},
 primaryClass = {astro-ph.CO},
       adsurl = {https://ui.adsabs.harvard.edu/abs/2012ApJ...753..114W}
}

@ARTICLE{Wuyts2013,
       author = {{Wuyts}, Stijn and {F{\"o}rster Schreiber}, Natascha M. and {Nelson}, Erica J. and {van Dokkum}, Pieter G. and {Brammer}, Gabe and {Chang}, Yu-Yen and {Faber}, Sandra M. and {Ferguson}, Henry C. and {Franx}, Marijn and {Fumagalli}, Mattia and {Genzel}, Reinhard and {Grogin}, Norman A. and {Kocevski}, Dale D. and {Koekemoer}, Anton M. and {Lundgren}, Britt and {Lutz}, Dieter and {McGrath}, Elizabeth J. and {Momcheva}, Ivelina and {Rosario}, David and {Skelton}, Rosalind E. and {Tacconi}, Linda J. and {van der Wel}, Arjen and {Whitaker}, Katherine E.},
        title = "{A CANDELS-3D-HST synergy: Resolved Star Formation Patterns at 0.7 < z < 1.5}",
      journal = {\apj},
         year = 2013,
        month = dec,
       volume = {779},
       number = {2},
          eid = {135},
        pages = {135},
          doi = {10.1088/0004-637X/779/2/135},
archivePrefix = {arXiv},
       eprint = {1310.5702},
 primaryClass = {astro-ph.CO},
       adsurl = {https://ui.adsabs.harvard.edu/abs/2013ApJ...779..135W}
}

@ARTICLE{Livermore2015,
       author = {{Livermore}, R.~C. and {Jones}, T.~A. and {Richard}, J. and {Bower}, R.~G. and {Swinbank}, A.~M. and {Yuan}, T.-T. and {Edge}, A.~C. and {Ellis}, R.~S. and {Kewley}, L.~J. and {Smail}, Ian and {Coppin}, K.~E.~K. and {Ebeling}, H.},
        title = "{Resolved spectroscopy of gravitationally lensed galaxies: global dynamics and star-forming clumps on {\ensuremath{\sim}}100 pc scales at 1 < z < 4}",
      journal = {\mnras},
         year = 2015,
        month = jun,
       volume = {450},
       number = {2},
        pages = {1812-1835},
          doi = {10.1093/mnras/stv686},
archivePrefix = {arXiv},
       eprint = {1503.07873},
 primaryClass = {astro-ph.GA},
       adsurl = {https://ui.adsabs.harvard.edu/abs/2015MNRAS.450.1812L}
}

@ARTICLE{Shibuya2016,
       author = {{Shibuya}, Takatoshi and {Ouchi}, Masami and {Kubo}, Mariko and {Harikane}, Yuichi},
        title = "{Morphologies of \raisebox{-0.5ex}\textasciitilde190,000 Galaxies at z = 0-10 Revealed with HST Legacy Data. II. Evolution of Clumpy Galaxies}",
      journal = {\apj},
         year = 2016,
        month = apr,
       volume = {821},
       number = {2},
          eid = {72},
        pages = {72},
          doi = {10.3847/0004-637X/821/2/72},
archivePrefix = {arXiv},
       eprint = {1511.07054},
 primaryClass = {astro-ph.GA},
       adsurl = {https://ui.adsabs.harvard.edu/abs/2016ApJ...821...72S}
}

@ARTICLE{Soto2017,
       author = {{Soto}, Emmaris and {de Mello}, Duilia F. and {Rafelski}, Marc and {Gardner}, Jonathan P. and {Teplitz}, Harry I. and {Koekemoer}, Anton M. and {Ravindranath}, Swara and {Grogin}, Norman A. and {Scarlata}, Claudia and {Kurczynski}, Peter and {Gawiser}, Eric},
        title = "{Physical Properties of Sub-galactic Clumps at 0.5 {\ensuremath{\leq}} Z {\ensuremath{\leq}} 1.5 in the UVUDF}",
      journal = {\apj},
         year = 2017,
        month = mar,
       volume = {837},
       number = {1},
          eid = {6},
        pages = {6},
          doi = {10.3847/1538-4357/aa5da3},
archivePrefix = {arXiv},
       eprint = {1702.03038},
 primaryClass = {astro-ph.GA},
       adsurl = {https://ui.adsabs.harvard.edu/abs/2017ApJ...837....6S}
}

@ARTICLE{Zanella2019,
       author = {{Zanella}, A. and {Le Floc'h}, E. and {Harrison}, C.~M. and {Daddi}, E. and {Bernhard}, E. and {Gobat}, R. and {Strazzullo}, V. and {Valentino}, F. and {Cibinel}, A. and {S{\'a}nchez Almeida}, J. and {Kohandel}, M. and {Fensch}, J. and {Behrendt}, M. and {Burkert}, A. and {Onodera}, M. and {Bournaud}, F. and {Scholtz}, J.},
        title = "{A contribution of star-forming clumps and accreting satellites to the mass assembly of z {\ensuremath{\sim}} 2 galaxies}",
      journal = {\mnras},
         year = 2019,
        month = oct,
       volume = {489},
       number = {2},
        pages = {2792-2818},
          doi = {10.1093/mnras/stz2099},
archivePrefix = {arXiv},
       eprint = {1907.12136},
 primaryClass = {astro-ph.GA},
       adsurl = {https://ui.adsabs.harvard.edu/abs/2019MNRAS.489.2792Z}
}

@ARTICLE{Mehta2021,
       author = {{Mehta}, Vihang and {Scarlata}, Claudia and {Fortson}, Lucy and {Dickinson}, Hugh and {Adams}, Dominic and {Chevallard}, Jacopo and {Charlot}, St{\'e}phane and {Beck}, Melanie and {Kruk}, Sandor and {Simmons}, Brooke},
        title = "{Investigating Clumpy Galaxies in the Sloan Digital Sky Survey Stripe 82 Using the Galaxy Zoo}",
      journal = {\apj},
         year = 2021,
        month = may,
       volume = {912},
       number = {1},
          eid = {49},
        pages = {49},
          doi = {10.3847/1538-4357/abed5b},
archivePrefix = {arXiv},
       eprint = {2011.01232},
 primaryClass = {astro-ph.GA},
       adsurl = {https://ui.adsabs.harvard.edu/abs/2021ApJ...912...49M}
}

@ARTICLE{Vanzella2021,
       author = {{Vanzella}, E. and {Caminha}, G.~B. and {Rosati}, P. and {Mercurio}, A. and {Castellano}, M. and {Meneghetti}, M. and {Grillo}, C. and {Sani}, E. and {Bergamini}, P. and {Calura}, F. and {Caputi}, K. and {Cristiani}, S. and {Cupani}, G. and {Fontana}, A. and {Gilli}, R. and {Grazian}, A. and {Gronke}, M. and {Mignoli}, M. and {Nonino}, M. and {Pentericci}, L. and {Tozzi}, P. and {Treu}, T. and {Balestra}, I. and {Dijkstra}, M.},
        title = "{The MUSE Deep Lensed Field on the Hubble Frontier Field MACS J0416. Star-forming complexes at cosmological distances}",
      journal = {\aap},
         year = 2021,
        month = feb,
       volume = {646},
          eid = {A57},
        pages = {A57},
          doi = {10.1051/0004-6361/202039466},
archivePrefix = {arXiv},
       eprint = {2009.08458},
 primaryClass = {astro-ph.GA},
       adsurl = {https://ui.adsabs.harvard.edu/abs/2021A&A...646A..57V}
}

@ARTICLE{Mestric2022,
       author = {{Me{\v{s}}tri{\'c}}, U. and {Vanzella}, E. and {Zanella}, A. and {Castellano}, M. and {Calura}, F. and {Rosati}, P. and {Bergamini}, P. and {Mercurio}, A. and {Meneghetti}, M. and {Grillo}, C. and {Caminha}, G.~B. and {Nonino}, M. and {Merlin}, E. and {Cupani}, G. and {Sani}, E.},
        title = "{Exploring the physical properties of lensed star-forming clumps at 2 {\ensuremath{\lesssim}} z {\ensuremath{\lesssim}} 6}",
      journal = {\mnras},
         year = 2022,
        month = nov,
       volume = {516},
       number = {3},
        pages = {3532-3555},
          doi = {10.1093/mnras/stac2309},
archivePrefix = {arXiv},
       eprint = {2202.09377},
 primaryClass = {astro-ph.GA},
       adsurl = {https://ui.adsabs.harvard.edu/abs/2022MNRAS.516.3532M}
}

@ARTICLE{Noguchi1998,
       author = {{Noguchi}, Masafumi},
        title = "{Clumpy star-forming regions as the origin of the peculiar morphology of high-redshift galaxies}",
      journal = {\nat},
         year = 1998,
        month = mar,
       volume = {392},
       number = {6673},
        pages = {253-256},
          doi = {10.1038/32596},
       adsurl = {https://ui.adsabs.harvard.edu/abs/1998Natur.392..253N}
}

@ARTICLE{Noguchi1999,
       author = {{Noguchi}, Masafumi},
        title = "{Early Evolution of Disk Galaxies: Formation of Bulges in Clumpy Young Galactic Disks}",
      journal = {\apj},
         year = 1999,
        month = mar,
       volume = {514},
       number = {1},
        pages = {77-95},
          doi = {10.1086/306932},
archivePrefix = {arXiv},
       eprint = {astro-ph/9806355},
 primaryClass = {astro-ph},
       adsurl = {https://ui.adsabs.harvard.edu/abs/1999ApJ...514...77N}
}

@ARTICLE{Immeli2004,
       author = {{Immeli}, A. and {Samland}, M. and {Gerhard}, O. and {Westera}, P.},
        title = "{Gas physics, disk fragmentation,  and bulge formation in young galaxies}",
      journal = {\aap},
         year = 2004,
        month = jan,
       volume = {413},
        pages = {547-561},
          doi = {10.1051/0004-6361:20034282},
archivePrefix = {arXiv},
       eprint = {astro-ph/0312139},
 primaryClass = {astro-ph},
       adsurl = {https://ui.adsabs.harvard.edu/abs/2004A&A...413..547I}
}

@ARTICLE{Immeli2004a,
       author = {{Immeli}, Andreas and {Samland}, Markus and {Westera}, Pieter and {Gerhard}, Ortwin},
        title = "{Subgalactic Clumps at High Redshift: A Fragmentation Origin?}",
      journal = {\apj},
         year = 2004,
        month = aug,
       volume = {611},
       number = {1},
        pages = {20-25},
          doi = {10.1086/422179},
archivePrefix = {arXiv},
       eprint = {astro-ph/0406135},
 primaryClass = {astro-ph},
       adsurl = {https://ui.adsabs.harvard.edu/abs/2004ApJ...611...20I}
}

@ARTICLE{Bournaud2009,
       author = {{Bournaud}, Fr{\'e}d{\'e}ric and {Elmegreen}, Bruce G.},
        title = "{Unstable Disks at High Redshift: Evidence for Smooth Accretion in Galaxy Formation}",
      journal = {\apjl},
         year = 2009,
        month = apr,
       volume = {694},
       number = {2},
        pages = {L158-L161},
          doi = {10.1088/0004-637X/694/2/L158},
archivePrefix = {arXiv},
       eprint = {0902.2806},
 primaryClass = {astro-ph.CO},
       adsurl = {https://ui.adsabs.harvard.edu/abs/2009ApJ...694L.158B}
}

@ARTICLE{Agertz2009,
       author = {{Agertz}, Oscar and {Teyssier}, Romain and {Moore}, Ben},
        title = "{Disc formation and the origin of clumpy galaxies at high redshift}",
      journal = {\mnras},
         year = 2009,
        month = jul,
       volume = {397},
       number = {1},
        pages = {L64-L68},
          doi = {10.1111/j.1745-3933.2009.00685.x},
archivePrefix = {arXiv},
       eprint = {0901.2536},
 primaryClass = {astro-ph.GA},
       adsurl = {https://ui.adsabs.harvard.edu/abs/2009MNRAS.397L..64A}
}

@ARTICLE{Ceverino2010,
       author = {{Ceverino}, Daniel and {Dekel}, Avishai and {Bournaud}, Frederic},
        title = "{High-redshift clumpy discs and bulges in cosmological simulations}",
      journal = {\mnras},
         year = 2010,
        month = jun,
       volume = {404},
       number = {4},
        pages = {2151-2169},
          doi = {10.1111/j.1365-2966.2010.16433.x},
archivePrefix = {arXiv},
       eprint = {0907.3271},
 primaryClass = {astro-ph.CO},
       adsurl = {https://ui.adsabs.harvard.edu/abs/2010MNRAS.404.2151C}
}

@ARTICLE{Romeo2010,
       author = {{Romeo}, Alessandro B. and {Burkert}, Andreas and {Agertz}, Oscar},
        title = "{A Toomre-like stability criterion for the clumpy and turbulent interstellar medium}",
      journal = {\mnras},
         year = 2010,
        month = sep,
       volume = {407},
       number = {2},
        pages = {1223-1230},
          doi = {10.1111/j.1365-2966.2010.16975.x},
archivePrefix = {arXiv},
       eprint = {1001.4732},
 primaryClass = {astro-ph.CO},
       adsurl = {https://ui.adsabs.harvard.edu/abs/2010MNRAS.407.1223R}
}

@ARTICLE{Romeo2014,
       author = {{Romeo}, Alessandro B. and {Agertz}, Oscar},
        title = "{Larson's scaling laws, and the gravitational instability of clumpy discs at high redshift}",
      journal = {\mnras},
         year = 2014,
        month = aug,
       volume = {442},
       number = {2},
        pages = {1230-1238},
          doi = {10.1093/mnras/stu954},
archivePrefix = {arXiv},
       eprint = {1403.0799},
 primaryClass = {astro-ph.GA},
       adsurl = {https://ui.adsabs.harvard.edu/abs/2014MNRAS.442.1230R}
}

@ARTICLE{Inoue2016,
       author = {{Inoue}, Shigeki and {Dekel}, Avishai and {Mandelker}, Nir and {Ceverino}, Daniel and {Bournaud}, Fr{\'e}d{\'e}ric and {Primack}, Joel},
        title = "{Non-linear violent disc instability with high Toomre's Q in high-redshift clumpy disc galaxies}",
      journal = {\mnras},
         year = 2016,
        month = feb,
       volume = {456},
       number = {2},
        pages = {2052-2069},
          doi = {10.1093/mnras/stv2793},
archivePrefix = {arXiv},
       eprint = {1510.07695},
 primaryClass = {astro-ph.GA},
       adsurl = {https://ui.adsabs.harvard.edu/abs/2016MNRAS.456.2052I}
}

@ARTICLE{Orr2024,
       author = {{Orr}, Matthew E. and {Rennehan}, Douglas},
        title = "{How the Cookie Crumbles: A Model for Star-forming Clumps in High-redshift Disk Galaxies}",
      journal = {arXiv e-prints},
         year = 2024,
        month = oct,
          eid = {arXiv:2410.23337},
        pages = {arXiv:2410.23337},
          doi = {10.48550/arXiv.2410.23337},
archivePrefix = {arXiv},
       eprint = {2410.23337},
 primaryClass = {astro-ph.GA},
       adsurl = {https://ui.adsabs.harvard.edu/abs/2024arXiv241023337O}
}

@ARTICLE{DiMatteo2008,
       author = {{Di Matteo}, P. and {Bournaud}, F. and {Martig}, M. and {Combes}, F. and {Melchior}, A.-L. and {Semelin}, B.},
        title = "{On the frequency, intensity, and duration of starburst episodes triggered by galaxy interactions and mergers}",
      journal = {\aap},
         year = 2008,
        month = dec,
       volume = {492},
       number = {1},
        pages = {31-49},
          doi = {10.1051/0004-6361:200809480},
archivePrefix = {arXiv},
       eprint = {0809.2592},
 primaryClass = {astro-ph},
       adsurl = {https://ui.adsabs.harvard.edu/abs/2008A&A...492...31D}
}

@ARTICLE{Renaud2015,
       author = {{Renaud}, Florent and {Bournaud}, Fr{\'e}d{\'e}ric and {Duc}, Pierre-Alain},
        title = "{A parsec-resolution simulation of the Antennae galaxies: formation of star clusters during the merger}",
      journal = {\mnras},
         year = 2015,
        month = jan,
       volume = {446},
       number = {2},
        pages = {2038-2054},
          doi = {10.1093/mnras/stu2208},
archivePrefix = {arXiv},
       eprint = {1410.5754},
 primaryClass = {astro-ph.GA},
       adsurl = {https://ui.adsabs.harvard.edu/abs/2015MNRAS.446.2038R}
}

@INPROCEEDINGS{Nakazato2024,
       author = {{Nakazato}, Yurina and {Ceverino}, Daniel and {Yoshida}, Naoki},
        title = "{FirstLight zoom-in simulations: Formation mechanism of [OIII]-bright clumps in high-redshift galaxies from z = 6-9}",
    booktitle = {EAS2024, European Astronomical Society Annual Meeting},
         year = 2024,
        month = jul,
          eid = {1261},
        pages = {1261},
       adsurl = {https://ui.adsabs.harvard.edu/abs/2024eas..conf.1261N}
}

@ARTICLE{Sattari2023,
       author = {{Sattari}, Zahra and {Mobasher}, Bahram and {Chartab}, Nima and {Kelson}, Daniel D. and {Teplitz}, Harry I. and {Rafelski}, Marc and {Grogin}, Norman A. and {Koekemoer}, Anton M. and {Wang}, Xin and {Windhorst}, Rogier A. and {Alavi}, Anahita and {Prichard}, Laura and {Sunnquist}, Ben and {Gardner}, Jonathan P. and {Gawiser}, Eric and {Hathi}, Nimish P. and {Hayes}, Matthew J. and {Ji}, Zhiyuan and {Mehta}, Vihang and {Robertson}, Brant E. and {Scarlata}, Claudia and {Yung}, L.~Y. Aaron and {Conselice}, Christopher J. and {Dai}, Y. Sophia and {Guo}, Yicheng and {Lucas}, Ray A. and {Martin}, Alec and {Ravindranath}, Swara},
        title = "{Fraction of Clumpy Star-forming Galaxies at 0.5 {\ensuremath{\leq}} z {\ensuremath{\leq}} 3 in UVCANDELS: Dependence on Stellar Mass and Environment}",
      journal = {\apj},
         year = 2023,
        month = jul,
       volume = {951},
       number = {2},
          eid = {147},
        pages = {147},
          doi = {10.3847/1538-4357/acd5d6},
archivePrefix = {arXiv},
       eprint = {2305.09021},
 primaryClass = {astro-ph.GA},
       adsurl = {https://ui.adsabs.harvard.edu/abs/2023ApJ...951..147S}
}

@ARTICLE{Claeyssens2023,
       author = {{Claeyssens}, Ad{\'e}la{\"\i}de and {Adamo}, Angela and {Richard}, Johan and {Mahler}, Guillaume and {Messa}, Matteo and {Dessauges-Zavadsky}, Miroslava},
        title = "{Star formation at the smallest scales: a JWST study of the clump populations in SMACS0723}",
      journal = {\mnras},
         year = 2023,
        month = apr,
       volume = {520},
       number = {2},
        pages = {2180-2203},
          doi = {10.1093/mnras/stac3791},
archivePrefix = {arXiv},
       eprint = {2208.10450},
 primaryClass = {astro-ph.GA},
       adsurl = {https://ui.adsabs.harvard.edu/abs/2023MNRAS.520.2180C}
}

@misc{Vega2026,
      title={The Fraction of Clumpy Galaxies in JADES over $2<z<9$}, 
      author={Alexander de la Vega and Bahram Mobasher and Zahra Sattari and Nima Chartab and Faezeh Manesh and Niloofar Sharei},
      year={2026},
      eprint={2508.14972},
      archivePrefix={arXiv},
      primaryClass={astro-ph.GA},
      url={https://arxiv.org/abs/2508.14972}, 
}

@ARTICLE{Fujimoto2025,
       author = {{Fujimoto}, S. and {Ouchi}, M. and {Kohno}, K. and {Valentino}, F. and {Gim{\'e}nez-Arteaga}, C. and {Brammer}, G.~B. and {Furtak}, L.~J. and {Kohandel}, M. and {Oguri}, M. and {Pallottini}, A. and {Richard}, J. and {Zitrin}, A. and {Bauer}, F.~E. and {Boylan-Kolchin}, M. and {Dessauges-Zavadsky}, M. and {Egami}, E. and {Finkelstein}, S.~L. and {Ma}, Z. and {Smail}, I. and {Watson}, D. and {Hutchison}, T.~A. and {Rigby}, J.~R. and {Welch}, B.~D. and {Ao}, Y. and {Bradley}, L.~D. and {Caminha}, G.~B. and {Caputi}, K.~I. and {Espada}, D. and {Endsley}, R. and {Fudamoto}, Y. and {Gonz{\'a}lez-L{\'o}pez}, J. and {Hatsukade}, B. and {Koekemoer}, A.~M. and {Kokorev}, V. and {Laporte}, N. and {Lee}, M. and {Magdis}, G.~E. and {Ono}, Y. and {Rizzo}, F. and {Shibuya}, T. and {Shimasaku}, K. and {Sun}, F. and {Toft}, S. and {Umehata}, H. and {Wang}, T. and {Yajima}, H.},
        title = "{Primordial rotating disk composed of at least 15 dense star-forming clumps at cosmic dawn}",
      journal = {Nature Astronomy},
         year = 2025,
        month = aug,
       volume = {9},
        pages = {1553-1567},
          doi = {10.1038/s41550-025-02592-w},
archivePrefix = {arXiv},
       eprint = {2402.18543},
 primaryClass = {astro-ph.GA},
       adsurl = {https://ui.adsabs.harvard.edu/abs/2025NatAs...9.1553F}
}

@ARTICLE{Mowla2024,
       author = {{Mowla}, Lamiya and {Iyer}, Kartheik and {Asada}, Yoshihisa and {Desprez}, Guillaume and {Tan}, Vivian Yun Yan and {Martis}, Nicholas and {Sarrouh}, Ghassan and {Strait}, Victoria and {Abraham}, Roberto and {Brada{\v{c}}}, Maru{\v{s}}a and {Brammer}, Gabriel and {Muzzin}, Adam and {Pacifici}, Camilla and {Ravindranath}, Swara and {Sawicki}, Marcin and {Willott}, Chris and {Estrada-Carpenter}, Vince and {Jahan}, Nusrath and {Noirot}, Ga{\"e}l and {Matharu}, Jasleen and {Rihtar{\v{s}}i{\v{c}}}, Gregor and {Zabl}, Johannes},
        title = "{Formation of a low-mass galaxy from star clusters in a 600-million-year-old Universe}",
      journal = {\nat},
         year = 2024,
        month = dec,
       volume = {636},
       number = {8042},
        pages = {332-336},
          doi = {10.1038/s41586-024-08293-0},
archivePrefix = {arXiv},
       eprint = {2402.08696},
 primaryClass = {astro-ph.GA},
       adsurl = {https://ui.adsabs.harvard.edu/abs/2024Natur.636..332M}
}

@ARTICLE{Bradley2025,
       author = {{Bradley}, Larry D. and {Adamo}, Angela and {Vanzella}, Eros and {Sharon}, Keren and {Brammer}, Gabriel and {Coe}, Dan and {Diego}, Jose M. and {Kokorev}, Vasily and {Mahler}, Guillaume and {Oguri}, Masamune and {Abdurro'uf} and {Bhatawdekar}, Rachana and {Christensen}, Lise and {Fujimoto}, Seiji and {Hashimoto}, Takuya and {Hsiao}, Tiger Y.-Y. and {Inoue}, Akio K. and {Jim{\'e}nez-Teja}, Yolanda and {Messa}, Matteo and {Norman}, Colin and {Ricotti}, Massimo and {Tamura}, Yoichi and {Windhorst}, Rogier A. and {Xu}, Xinfeng and {Zitrin}, Adi},
        title = "{Unveiling the Cosmic Gems Arc at z {\ensuremath{\sim}} 10 with JWST NIRCam}",
      journal = {\apj},
         year = 2025,
        month = sep,
       volume = {991},
       number = {1},
          eid = {32},
        pages = {32},
          doi = {10.3847/1538-4357/adf638},
archivePrefix = {arXiv},
       eprint = {2404.10770},
 primaryClass = {astro-ph.GA},
       adsurl = {https://ui.adsabs.harvard.edu/abs/2025ApJ...991...32B}
}

@ARTICLE{Tacchella2023,
       author = {{Tacchella}, Sandro and {Johnson}, Benjamin D. and {Robertson}, Brant E. and {Carniani}, Stefano and {D'Eugenio}, Francesco and {Kumari}, Nimisha and {Maiolino}, Roberto and {Nelson}, Erica J. and {Suess}, Katherine A. and {{\"U}bler}, Hannah and {Williams}, Christina C. and {Adebusola}, Alabi and {Alberts}, Stacey and {Arribas}, Santiago and {Bhatawdekar}, Rachana and {Bonaventura}, Nina and {Bowler}, Rebecca A.~A. and {Bunker}, Andrew J. and {Cameron}, Alex J. and {Curti}, Mirko and {Egami}, Eiichi and {Eisenstein}, Daniel J. and {Frye}, Brenda and {Hainline}, Kevin and {Helton}, Jakob M. and {Ji}, Zhiyuan and {Looser}, Tobias J. and {Lyu}, Jianwei and {Perna}, Michele and {Rawle}, Timothy and {Rieke}, George and {Rieke}, Marcia and {Saxena}, Aayush and {Sandles}, Lester and {Shivaei}, Irene and {Simmonds}, Charlotte and {Sun}, Fengwu and {Willmer}, Christopher N.~A. and {Willott}, Chris J. and {Witstok}, Joris},
        title = "{JWST NIRCam + NIRSpec: interstellar medium and stellar populations of young galaxies with rising star formation and evolving gas reservoirs}",
      journal = {\mnras},
         year = 2023,
        month = jul,
       volume = {522},
       number = {4},
        pages = {6236-6249},
          doi = {10.1093/mnras/stad1408},
archivePrefix = {arXiv},
       eprint = {2208.03281},
 primaryClass = {astro-ph.GA},
       adsurl = {https://ui.adsabs.harvard.edu/abs/2023MNRAS.522.6236T}
}

@misc{Hainline2024,
      title={The Cosmos in its Infancy: JADES Galaxy Candidates at z > 8 in GOODS-S and GOODS-N}, 
      author={Kevin N. Hainline and Benjamin D. Johnson and Brant Robertson and Sandro Tacchella and Jakob M. Helton and Fengwu Sun and Daniel J. Eisenstein and Charlotte Simmonds and Michael W. Topping and Lily Whitler and Christopher N. A. Willmer and Marcia Rieke and Katherine A. Suess and Raphael E. Hviding and Alex J. Cameron and Stacey Alberts and William M. Baker and Rachana Bhatawdekar and Kristan Boyett and Andrew J. Bunker and Stefano Carniani and Stephane Charlot and Zuyi Chen and Mirko Curti and Emma Curtis-Lake and Francesco D'Eugenio and Eiichi Egami and Ryan Endsley and Ryan Hausen and Zhiyuan Ji and Tobias J. Looser and Jianwei Lyu and Roberto Maiolino and Erica Nelson and David Puskas and Tim Rawle and Lester Sandles and Aayush Saxena and Renske Smit and Daniel P. Stark and Christina C. Williams and Chris Willott and Joris Witstok},
      year={2024},
      eprint={2306.02468},
      archivePrefix={arXiv},
      primaryClass={astro-ph.GA},
      url={https://arxiv.org/abs/2306.02468}, 
}

@ARTICLE{Chen2023,
       author = {{Chen}, Zuyi and {Stark}, Daniel P. and {Endsley}, Ryan and {Topping}, Michael and {Whitler}, Lily and {Charlot}, St{\'e}phane},
        title = "{JWST/NIRCam observations of stars and H II regions in z ≃ 6-8 galaxies: properties of star-forming complexes on 150 pc scales}",
      journal = {\mnras},
         year = 2023,
        month = feb,
       volume = {518},
       number = {4},
        pages = {5607-5619},
          doi = {10.1093/mnras/stac3476},
archivePrefix = {arXiv},
       eprint = {2207.12657},
 primaryClass = {astro-ph.GA},
       adsurl = {https://ui.adsabs.harvard.edu/abs/2023MNRAS.518.5607C}
}

@ARTICLE{Costantin2023,
       author = {{Costantin}, Luca and {P{\'e}rez-Gonz{\'a}lez}, Pablo G. and {Guo}, Yuchen and {Buttitta}, Chiara and {Jogee}, Shardha and {Bagley}, Micaela B. and {Barro}, Guillermo and {Kartaltepe}, Jeyhan S. and {Koekemoer}, Anton M. and {Cabello}, Cristina and {Corsini}, Enrico Maria and {M{\'e}ndez-Abreu}, Jairo and {de la Vega}, Alexander and {Iyer}, Kartheik G. and {Bisigello}, Laura and {Cheng}, Yingjie and {Morelli}, Lorenzo and {Arrabal Haro}, Pablo and {Buitrago}, Fernando and {Cooper}, M.~C. and {Dekel}, Avishai and {Dickinson}, Mark and {Finkelstein}, Steven L. and {Giavalisco}, Mauro and {Holwerda}, Benne W. and {Huertas-Company}, Marc and {Lucas}, Ray A. and {Papovich}, Casey and {Pirzkal}, Nor and {Seill{\'e}}, Lise-Marie and {Vega-Ferrero}, Jes{\'u}s and {Wuyts}, Stijn and {Yung}, L.~Y. Aaron},
        title = "{A Milky Way-like barred spiral galaxy at a redshift of 3}",
      journal = {\nat},
         year = 2023,
        month = nov,
       volume = {623},
       number = {7987},
        pages = {499-501},
          doi = {10.1038/s41586-023-06636-x},
archivePrefix = {arXiv},
       eprint = {2311.04283},
 primaryClass = {astro-ph.GA},
       adsurl = {https://ui.adsabs.harvard.edu/abs/2023Natur.623..499C}
}

@ARTICLE{Lee2024,
       author = {{Lee}, Jeong Hwan and {Park}, Changbom and {Hwang}, Ho Seong and {Kwon}, Minseong},
        title = "{Morphology of Galaxies in JWST Fields: Initial Distribution and Evolution of Galaxy Morphology}",
      journal = {\apj},
         year = 2024,
        month = may,
       volume = {966},
       number = {1},
          eid = {113},
        pages = {113},
          doi = {10.3847/1538-4357/ad3448},
archivePrefix = {arXiv},
       eprint = {2312.04899},
 primaryClass = {astro-ph.GA},
       adsurl = {https://ui.adsabs.harvard.edu/abs/2024ApJ...966..113L}
}

@article{Wang2025,
   title={A giant disk galaxy two billion years after the Big Bang},
   volume={9},
   ISSN={2397-3366},
   url={http://dx.doi.org/10.1038/s41550-025-02500-2},
   DOI={10.1038/s41550-025-02500-2},
   number={5},
   journal={Nature Astronomy},
   publisher={Springer Science and Business Media LLC},
   author={Wang, Weichen and Cantalupo, Sebastiano and Pensabene, Antonio and Galbiati, Marta and Travascio, Andrea and Steidel, Charles C. and Maseda, Michael V. and Pezzulli, Gabriele and de Beer, Stephanie and Fossati, Matteo and Fumagalli, Michele and Gallego, Sofia G. and Lazeyras, Titouan and Mackenzie, Ruari and Matthee, Jorryt and Nanayakkara, Themiya and Quadri, Giada},
   year={2025},
   month=Mar, pages={710–719} }



\appendix

\section{Power spectra for individual galaxies}\label{apdx:ind}

\begin{figure}
    \centering
    \includegraphics[width=0.95\linewidth]{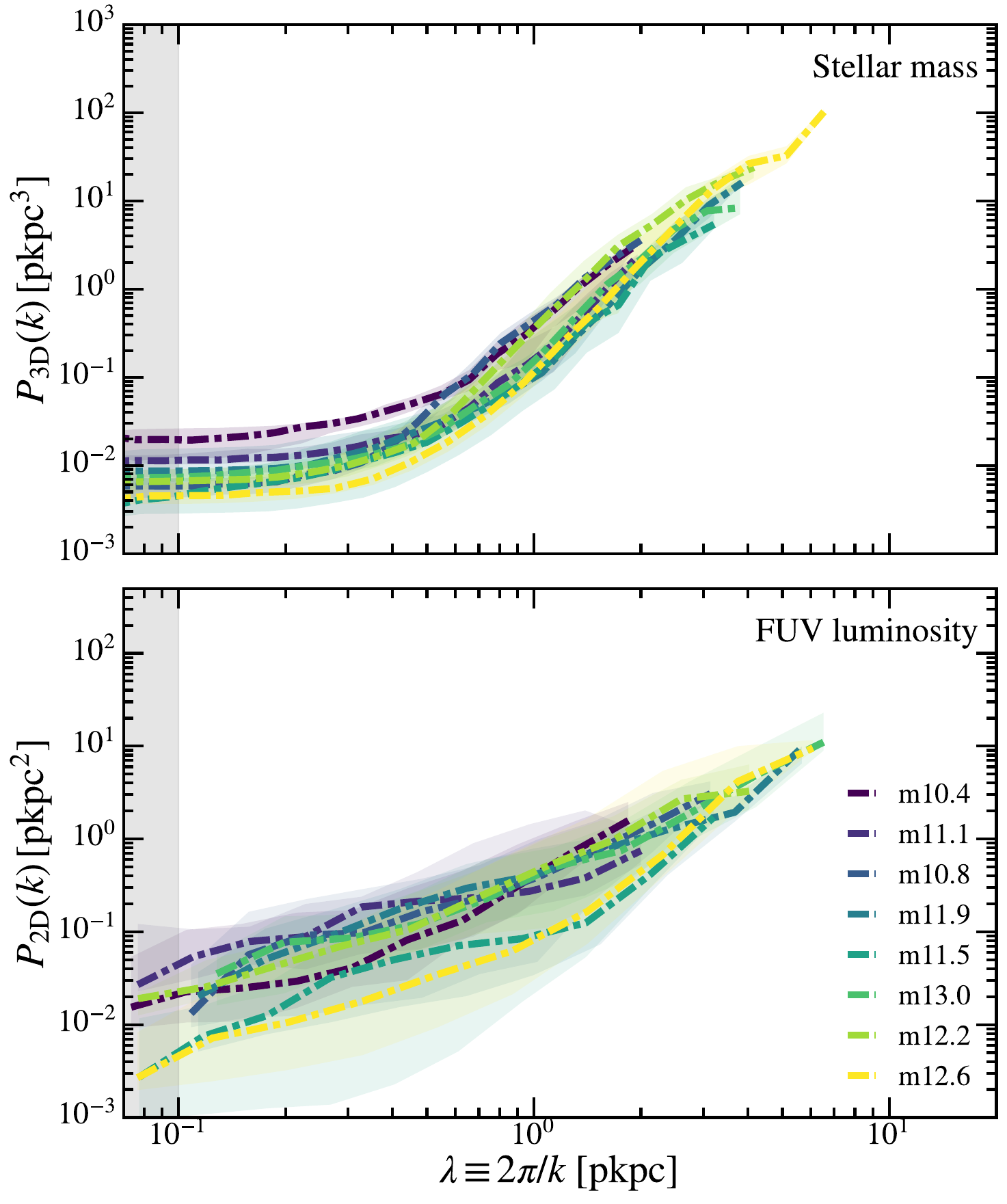}
    \caption{ 
    Power spectra of eight \thesanzoom galaxies at $z\simeq6$. Solid lines show median spectra over snapshots spanning one dynamical time, and shaded regions denote the $16{\mathrm{th}}$--$84{\mathrm{th}}$ percentile temporal variation. Colours are ordered by the median stellar mass of each galaxy over the same interval. The upper panel shows the three-dimensional stellar mass power spectrum, and the lower panel shows the projected FUV luminosity power spectrum. Galaxy-to-galaxy variation is modest, although lower-mass systems have systematically enhanced small-scale power, particularly in stellar mass. The FUV spectra vary more strongly with time but follow the same trend.
}
    \label{fig:gals}
\end{figure}

In the main text, we use median power spectra to reduce the stochastic variation associated with individual galaxies and snapshots. Here, we show the spectra of individual galaxies at $z\simeq6$ as an illustrative example. Figure~\ref{fig:gals} presents the median spectrum of each galaxy, computed over snapshots within one dynamical time, with colours ordered by the galaxy's median stellar mass over the same interval. The overall spectral shape is similar across galaxies, indicating that the galaxy-to-galaxy variance is modest. Lower-mass systems nevertheless have a relatively higher small-scale plateau in the stellar mass power spectrum. This behaviour is expected from particle discreteness. In the shot-noise-dominated regime, the rms fractional fluctuation scales as $N^{-1/2}$, while the corresponding white-noise power floor scales as $N^{-1}$ at fixed volume and normalization, where $N$ is the number of stellar particles. Lower-mass galaxies, which contain fewer stellar particles, therefore exhibit a higher small-scale power floor. The emission maps vary more strongly across snapshots, but retain the same qualitative tendency for lower-mass systems to appear more clumpy.

\begin{figure}
    \centering
    \includegraphics[width=0.95\linewidth]{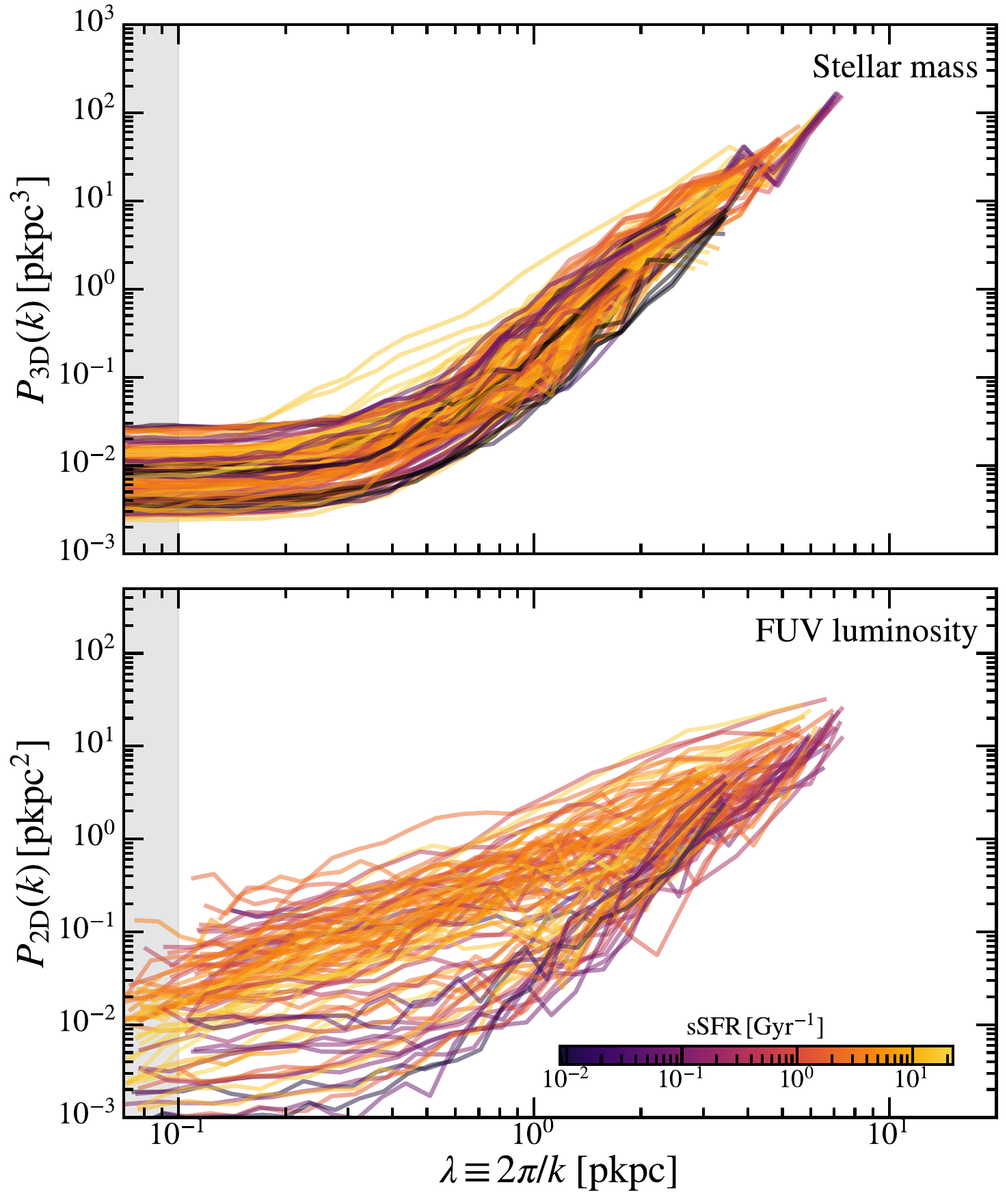}
    \caption{As in Figure~\ref{fig:gals}, but for individual snapshots at $z\simeq6$, colour-coded by sSFR. Individual snapshots have noisy, non-smooth spectral features that probably trace short-timescale fluctuations associated with transient clumpy structures. These features are largely absent from the dynamical-time-averaged spectra in Figure~\ref{fig:gals}. Snapshots with higher sSFR tend to have enhanced small-scale power, particularly in FUV emission.}
    \label{fig:ssfr}
\end{figure}

Figure~\ref{fig:ssfr} further presents the spectra of individual snapshots. Unlike the dynamical-time-averaged spectra, individual snapshots show irregular, non-smooth spectral features. These features arise from short-timescale fluctuations associated with transient clumpy structures and are smoothed out once the spectra are averaged over a dynamical time. We colour-code the snapshots by their specific star formation rate (sSFR). The resulting trend agrees with Figure~\ref{fig:bursty}. Galaxies in more active star-forming phases tend to have enhanced small-scale power and appear more clumpy, particularly in FUV emission.

\section{Impact of numerical resolution}\label{apdx:res}
\begin{figure}
    \centering
    \includegraphics[width=0.95\linewidth]{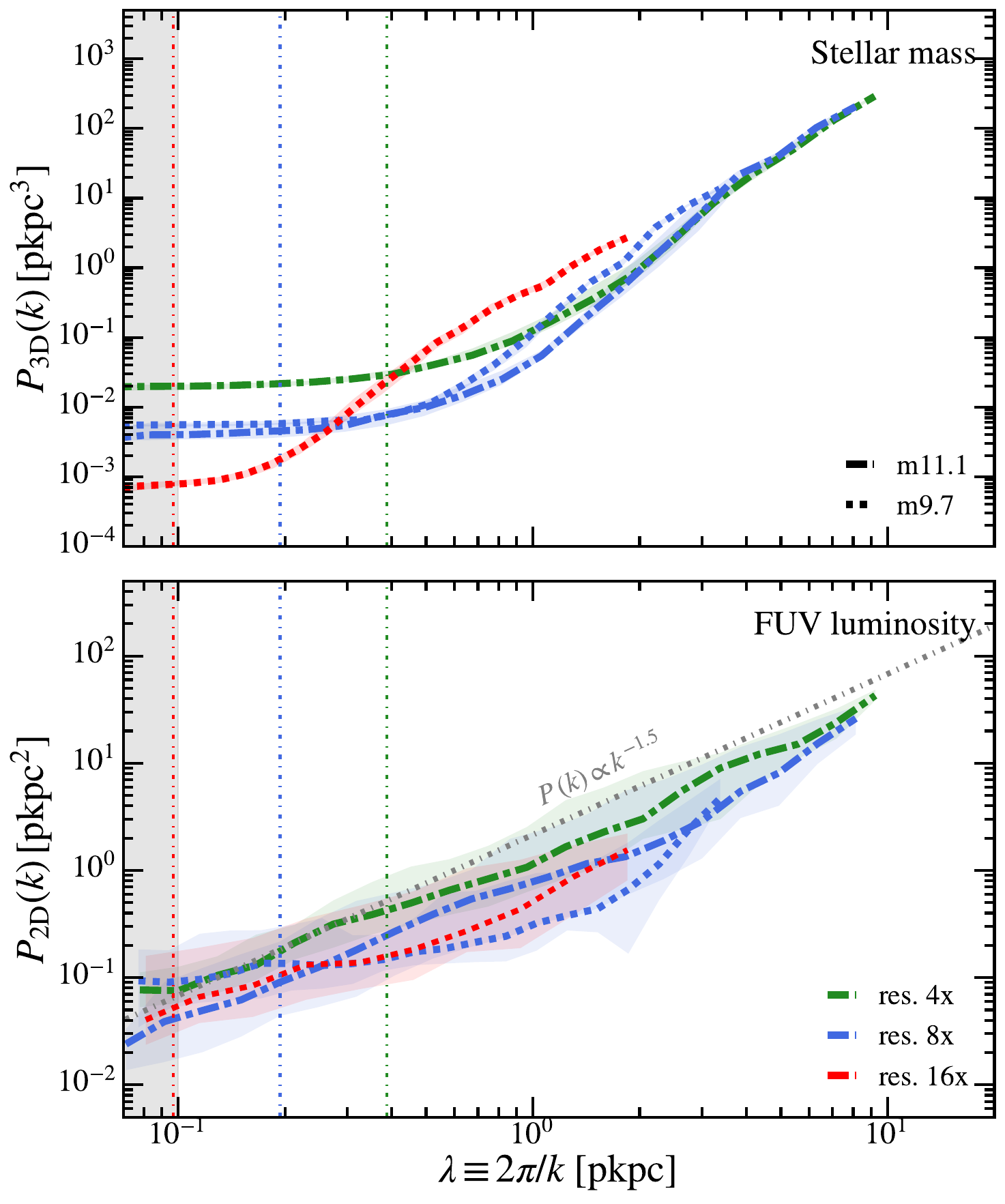}
    \caption{As in Figure~\ref{fig:model}, but for different \thesanzoom resolution levels. Vertical dashed lines mark the numerical gravitational softening scale, $2.8\epsilon_\ast$, for the selected galaxy sample at $z\simeq3$. The mass spectra at all three resolutions follow the trend in Figure~\ref{fig:FFT}, saturating near the resolution limit. The light spectra instead agree well across resolutions, showing that their power-law structure is not a numerical artifact.}
    \label{fig:resolution}
    \label{lastpage}
\end{figure}

In this section, we examine the effect of numerical resolution on the power spectra of both mass and light. Because higher-resolution runs are available only for lower-mass haloes, we select the most massive main-target galaxies that have been simulated at multiple resolution levels at $z=3$. Specifically, we compare ``m11.1'' between the $4\times$ and $8\times$ runs, and ``m9.7'' between the $4\times$ and $16\times$ runs.

Figure~\ref{fig:resolution} shows the resolution dependence of the mass and light power spectra. At all three resolution levels, the mass spectra decline smoothly and rapidly towards small scales before reaching a plateau near their respective softening lengths. This behaviour confirms that the relatively large stellar softening length in the simulations can erase long-lived compact stellar structures. By contrast, the light spectra agree well across all three resolutions. This agreement probably arises because FUV emission more closely traces the dense star-forming gas, whose effective gas-cell softening length is smaller than the Nyquist scale probed here, and therefore avoids the artificial plateau seen in the stellar mass spectra.



\bsp	
\end{document}